\pdfoutput=1
\documentclass[12pt,twoside,journal,draftclsnofoot,onecolumn]{IEEEtran}

\usepackage{amsmath,amsfonts,bm,amssymb,psfrag,ifthen,color,subfigure}
\usepackage{mathrsfs}
\usepackage[justification=centering,font=footnotesize]{caption}
\usepackage{todonotes}
\allowdisplaybreaks[4]

\usepackage[dvips]{epsfig}
\usepackage[dvips]{graphicx}
\usepackage{amsmath,amsfonts,bm,amssymb,psfrag,ifthen,color,subfigure}
\usepackage{algpseudocode}
\usepackage{algorithm}
\usepackage{algorithmicx}
\usepackage{ifthen}
\usepackage{setspace}
\usepackage{cite}
\usepackage{url}
\usepackage{longtable}
\usepackage{stfloats}  
\usepackage{graphics,booktabs,color,subfigure}
\usepackage{float}
\usepackage{diagbox}

\usepackage[dvips]{epsfig}
\usepackage[dvips]{graphicx}
\usepackage{amsmath,amsfonts,bm,amssymb,psfrag,ifthen,color,subfigure,amsthm}
\usepackage{algpseudocode}
\usepackage{algorithm}
\usepackage{algorithmicx}
\usepackage{setspace}
\usepackage{cite}
\usepackage{url}
\usepackage{longtable}
\usepackage{stfloats}  
\usepackage{graphics,booktabs,color,subfigure}
\usepackage{float}
\usepackage{diagbox}
\usepackage{multirow}

\allowdisplaybreaks[4]

\newcommand{\tabincell}[2]{\begin{tabular}{@{}#1@{}}#2\end{tabular}}

\newtheorem{Lemma}{Lemma}

\usepackage{graphicx}

 \usepackage{graphicx}
\usepackage{todonotes} 

\begin{document}
\title{Features Disentangled   Semantic
   Broadcast  Communication Networks}
\author{Shuai Ma,  Weining Qiao,   Youlong Wu,  Hang Li, Guangming Shi,~\IEEEmembership{Fellow,~IEEE}, Dahua Gao, Yuanming Shi,    Shiyin Li, and Naofal Al-Dhahir,~\IEEEmembership{Fellow,~IEEE}
			
			\thanks{Shuai Ma is with
				Pengcheng Laboratory, Shenzhen, 518066, China (e-mail: mash01@pcl.ac.cn).}

		}
		
\maketitle
\begin{abstract}

Single-user semantic communications have attracted extensive research recently, but multi-user semantic broadcast communication (BC) is still in its infancy.
        In this paper, we propose a  practical robust features-disentangled  multi-user semantic BC framework,
        where the transmitter includes a feature selection module and each user has a feature
completion module.
         Instead of broadcasting all extracted features, the semantic encoder  extracts the disentangled semantic features, and then only  the users' intended semantic features are selected   for broadcasting,   which can further improve the transmission efficiency. Within this framework,
         we further investigate two information-theoretic metrics, including the ultimate compression rate under both the distortion and  perception constraints, and the achievable rate region of the semantic BC.
   Furthermore,  to realize  the proposed semantic BC framework,
we    design a lightweight robust
    semantic BC network by exploiting   a supervised autoencoder (AE), which can  controllably disentangle sematic features.
 Moreover, we   design the first   hardware proof-of-concept prototype  of   the  semantic BC network,
 where the proposed  semantic BC network can be implemented in real time.
    Simulations and experiments demonstrate that the proposed
         robust   semantic   BC network can significantly improve transmission
      efficiency.

\end{abstract}

\IEEEpeerreviewmaketitle
  \section{Introduction}

  Due to the increasing   quality of service  (QoS) demands from the diverse  Internet of Things (IoT) devices, next-generation 6G communication
networks face significant challenges, such as
 huge volumes of data traffic,   ultra-high
 speed,   and extreme low latency requirements, which are driven by the applications of   holographic communications and
extremely-high-definition video transmissions\cite{Letaief_ICM_2019,Zhang_Engineering_2022,Kountouris_CM_2021}.
For
example, in the scenario of 8K surveillance video analysis,
the generated data size is about 12 Terabytes per hour \cite{Furuya_2009}.
Collecting such heavy workloads needs nearly three hours at a 5G transmission speed of 1 Gbps.
Cisco predicts that by  the end of 2023,   the number of
mobile  Internet-enabled devices will reach 29.3 billion, from 18.4 billion
in 2018\cite{Cisco_2021}, and the wireless data
traffic is estimated by the  international telecommunication union (ITU) to reach 4394 EB in 2030 \cite{Union_2015}.
 Under the   hardware cost  and   energy consumption limitations, such explosive demand will
gradually exceed the   capabilities of 5G networks \cite{Bao_ICST_2022,Niu_arXiv_2022}.

To extend 5G capabilities, semantic
communications,  which exploits computing power at the transceivers to  alleviate the cost of  transmission  resources, have emerged as a promising   key  6G technology \cite{Sana_CCNC_2022,Shi_CM_2021,Luo_WC_2022,Bao_INSW_2011,Yener_TCCN_2018}.
 Specifically, in contrast to conventional bit-level communication systems,  semantic communications   extract  and transmit
only task-relevant information,  and thus significantly alleviate the data traffic burden over the communication networks.  The key challenges in semantic
communications are how to precisely extract and efficiently deliver the task-relevant information to the destinations.

\subsection{Related works}

  Fortunately, recent advancements of
artificial intelligence (AI)  pave the
way to develop semantic communications for future wireless networks. Specifically, semantic communications have attracted intensive research efforts in text \cite{Bao_INSW_2011,Farsad_ICASSP_2018,Xie_TSP_2021,Jiang_arXiv_2021,Sana_CCNC_2022,Lu_arxiv_2021}, speech/audio \cite{Weng_JSAC_2021,Tong_GLOBECOM_2021,Shi_speech_2021}, video\cite{Jiang_arxiv_2022,Tung_arxiv_2022},
and image transmission \cite{Kurka_TWC_2021,Jankowski_ISAC_2021,Shao_JSAC_2022,Kang_arXiv_2021,Hu_arXiv_2022,Huang_GLOBECOM_2021,Bourtsoulatze_TCCN_2019,Xu_TCSVT_2021,Kurika_JSAC_2020,Yang_TCCN_2022,Yang_arXiv_2021}.
Specifically,  a joint semantic-channel
coding (JSCC) system  was  developed in \cite{Xie_TSP_2021} to minimize
the semantic errors  for text transmission.
By combining   semantic-channel
coding, a hybrid automatic repeat request (HARQ) was proposed in\cite{Jiang_arXiv_2021} to
improve  sentence semantic transmission efficiency.
An adaptive
end-to-end semantic system was designed in \cite{Sana_CCNC_2022} to maximize text transmission accuracy. A
reinforcement
learning (RL)  based semantic  learning scheme was designed   in  \cite{Lu_arxiv_2021} to   maximize
the semantic similarity of transmitted messages.
Besides, for semantic-aware  speech   transmission,
an attention mechanism-powered module was explored in \cite{Weng_JSAC_2021} to enhance robustness against the channel variations.
A convolutional  neural network (CNN) based  federated
learning model was designed in \cite{Tong_GLOBECOM_2021} for  multi-user audio semantic communication networks.
In \cite{Shi_speech_2021},
an    understanding-based automatic speech recognition architecture was developed for speech transmission with   high semantic fidelity.
For semantic-aware  video   transmission,
 an incremental redundancy hybrid automatic repeat request
(IR-HARQ) framework was proposed in \cite{Jiang_arxiv_2022} for   wicked channels in semantic video conferencing.
In \cite{Tung_arxiv_2022}, a deep learning-based JSCC solution for
wireless video transmission, called DeepWiVe, was proposed, where the bandwidth allocation is optimized though RL.

For image   semantic transmission,   a
deep-learning-based multiple-description JSCC scheme with adaptive bandwidth was proposed in \cite{Kurka_TWC_2021}.
To reduce the
transmission bandwidth requirement, a retrieval-oriented  image compression scheme was investigated \cite{Jankowski_ISAC_2021}  for the edge
network.
Based on the information bottleneck (IB)
principle,  a variable-length
  semantic feature encoding method was designed in   \cite{Shao_JSAC_2022} for   image classification.
By leveraging deep reinforcement learning (DRL),
a semantic image transmission scheme was designed in \cite{Kang_arXiv_2021} for scene classification.
 By utilizing the masked autoencoder,  a robust semantic
communication system was proposed   in \cite{Hu_arXiv_2022}   to combat  the   channel noise.
  A generative adversarial
networks (GANs)-based  semantic coding scheme was investigated in \cite{Huang_GLOBECOM_2021} for low bit-rate image  transmission.
By leveraging  CNNs  as
 the
encoder and the decoder,  a JSCC scheme  architecture was developed in \cite{Bourtsoulatze_TCCN_2019} for
   wireless image transmission.
 By employing attention mechanisms,  a JSCC method was designed to  automatically
 adjust  the image compression ratio  based    channel SNRs\cite{Xu_TCSVT_2021}.
 A deep neural network (DNN)-based JSCC
coding method that exploits the channel output feedback to improve the reconstruction image quality was investigated in \cite{Kurika_JSAC_2020}.
By  mapping the compressed
  images to OFDM samples, a CNNs based JSCC method  was designed in \cite{Yang_TCCN_2022} to combat multi-path fading.
  By employing the Gumbel-Softmax method, a  DNN based
 JSCC  structure was proposed in \cite{Yang_arXiv_2021} to
 dynamically assign  the rate based on the channel SNR and image content.

\subsection{Motivations and contributions}

It is worth  pointing out that most  existing works apply DL techniques in the transmission design. The main drawback of the DL-based models lies in the uninterpretability of the operation. Specifically, the extracted semantic features from the source data are encoded and coupled together, which are unexplainable (hidden) representations. Such a black-box issue prevents the application of the specific semantic features.
Meanwhile, due to the  hidden representations,
 the unintended semantic features  may also be transmitted to the receiver, which   may reduce the transmission efficiency.

  Another issue is that current research studies mainly focus on single-user point-to-point communication scenarios, while there are few studies on multi-user   semantic broadcast communications (BC). In \cite{Ding_ICASSP_2021},
an autoencoder-based   deep JSCC scheme was proposed  for
multi-user broadcast image transmission, where all the receivers wish to recover the same source image with the loss of total mean square error distortion. In fact, for   multi-user semantic BC,   the users may be interested in different semantic information, and the knowledge base  at the users could be also different. Thus, to enable efficient multi-user semantic BCs, one should exploit the variety of users' intended semantic information,
the BC channels properties, and  the assistant information at the transmitter and receivers (e.g., knowledge base).

In this paper, we   propose a robust features-disentangled     semantic   BC   framework, which    incorporates the well-established   bit-level communication system.  Furthermore,  a   lightweight   robust semantic BC network and the corresponding hardware proof-of-concept prototype are designed and devloped.
The main contributions of this paper are summarized as follows:
\begin{itemize}

\item To simplify implementation,
we propose a  features-disentangled broadcast semantic BC  framework,   which  is compatible with existing well-established communications systems.
   The advantages of our proposed semantic BC
    framework are three-fold: i)
 Instead of broadcasting all extracted features, the extracted semantic features
are firstly disentangled, and then only  the users' intended semantic features are selected   for broadcasting, while the unintended   features will not be transmitted, which can further improve the transmission efficiency;
ii) The semantic encoder not only
    compresses the input data to the  low-dimensional  features  as a source encoder,
   but also   improves robustness of the extracted  semantic features for both channel fading and channel noise  as  a channel encoder;
 iii) it    simultaneously
      takes advantage of the high transmission efficiency of semantic communications and the practical standards and hardware of the existing  well-designed communication networks.

\item Within the proposed semantic BC framework, we further investigate two information-theoretic metrics: the ultimate compression rate with
       both the
  distortion and   perception constraints \cite{Blau_ICML_2019}, and the achievable rate region of the semantic BC.  Specifically,    under both squared error distortion and the perception constraints,   we   propose the optimal distortion
allocation scheme for
multi-source data compression.
Moreover,  since  the semantic channel noise follows a non-Gaussian
distribution, the classical Shannon capacity   results   can not be directly applied for semantic BC channels.
     To quantify the semantic  information transmission, we derive both inner and outer bounds  for the achievable rate region semantic BC channels, which are   tight when the semantic channel noise    tends to the Gaussian
distribution.

\item  To realize  the proposed semantic BC framework,
we    design a lightweight robust
    semantic BC network by exploiting  a supervised autoencoder (AE), which can  controllably disentangle sematic features.
Specifically, motivated by the  group supervised learning strategy \cite{Ge_ICLR_2021},
we  jointly train the semantic BC encoder and multiple semantic decoders  in   three steps:
    self reconstruction, common semantic features exchange,   and  different semantic features exchange.    Moreover, to
 enhance  robust transmission, the channel  fading  and random channel noise  are considered during the proposed semantic BC network training.

\item Finally, we   design  a   hardware proof-of-concept prototype for   the  semantic BC network
  by utilizing  the  portable Jetson Nano B01
processors,
which include    one transmitter and two semantic
mobile users.  To  our best knowledge, this is the first prototype for semantic BC networks.
Specifically,  the  proposed    semantic BC network is implemented  based on  the  designed prototype platform in real time, and   the extracted intended
semantic   features are broadcasted  via Wi-Fi.
 The prototype experiments demonstrate  that  the proposed      semantic   BC network can
   significantly reduce transmission time compared to   existing benchmarks.

\end{itemize}

	\begin{table}[H]
	\caption{Key Notations   and   Meanings}
	\label{tablepar}
	\centering
	\begin{tabular}{|c|l|}
		\hline
		\rule{0pt}{8pt}Variables  & Meanings   \\ \hline
		\rule{0pt}{7.5pt}$K$ &  \tabincell{c}{Total number of semantic users
		} \\ \hline
		
		\rule{0pt}{7.5pt}${\mathcal{L}} \buildrel \Delta \over = \left\{ {1,...,L} \right\} $ &  \tabincell{c}{Semantic feature index set}

		\\ \hline
		
		\rule{0pt}{7.5pt}$Z = {\left\{ {{z_l}} \right\}_{l \in {\mathcal{L}}}}$ &  \tabincell{c}{Set of disentangled semantic    features
		}
		
		\\ \hline
		\rule{0pt}{7.5pt} $ {{{\mathcal{L}}_i}}  \subseteq {\mathcal{L}}$ &  \tabincell{c}{Interested feature index set of User $i$}\\ \hline

		\rule{0pt}{7.5pt}${X_{i,{\rm{s}}}} = {\left\{ {{z_l}} \right\}_{l \in {{\mathcal{L}_i}}}}$ &  \tabincell{c}{Selected  semantic features of User $i$}\\ \hline
		\rule{0pt}{7.5pt}${Y_{i,{\rm{s}}}} = {\left\{ {{{\widehat z}_l}} \right\}_{l \in {{\cal L}_{{{i}}}}}}$ &  \tabincell{c}{Estimated  semantic features of User $i$ }\\ \hline
		\rule{0pt}{7.5pt}$\widehat Z_i$  &  \tabincell{c}{Completed semantic features of User $i$}\\ \hline
	\end{tabular}
\end{table}

	The rest of this paper is organized as follows. 
 The
 features-disentangled semantic   BC network  framework is presented in Section II.
Section III  provides the information-theoretic metrics of a semantic BC network.
  In Section IV,    we  propose  a feasible  robust   semantic BC network.
 In Section V, we present the   semantic
BC network prototype design and implementation.
  Experimental results and analysis are presented in Section VI. Finally,
 Section VII provides the conclusions. Table I presents the
meaning of the key notations used in this paper.

\begin{figure*}[htbp]
	\centering
	\includegraphics[width=0.9\textwidth]{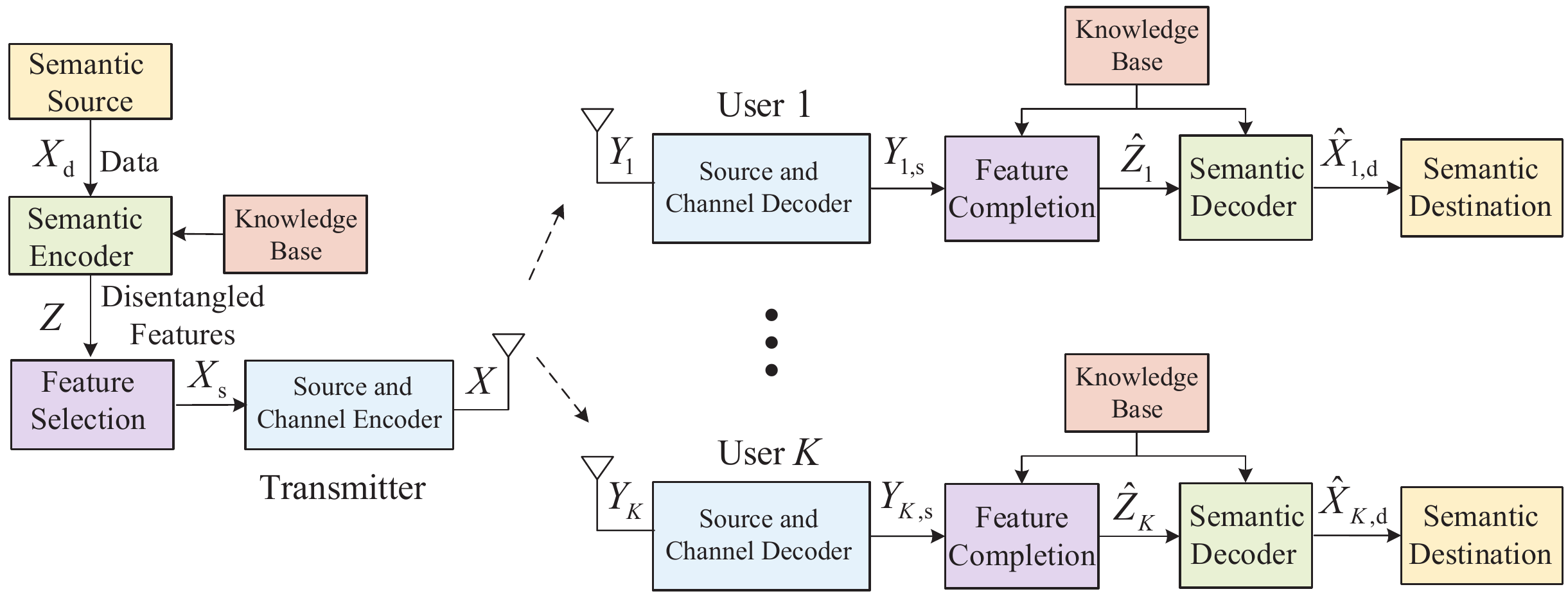}
	\caption{Features disentangled broadcast semantic communications  framework }
	\label{PTP_SM}
\end{figure*}

 \section{Features Disentangled Semantic BC Framework  }

In order to design a practical semantic communication network,
we propose a  features-disentangled broadcast semantic BC  framework, as shown in Fig.~1, which  is compatible with existing 5G communication systems.
Specifically,    one semantic transmitter  broadcasts the disentangled features to $K$  semantic users.
The proposed semantic broadcast framework can  simultaneously
      take advantage of the high transmission efficiency of semantic communications and the practical standards and hardware of    5G communications systems.
In the following, we will introduce the modules of the proposed semantic broadcast framework   in details.

  \subsubsection{Semantic Source} The semantic source produces original  data  ${X_{\rm{d}}}$, which
 generally     includes both the intended  semantic features and some unintended semantic features.
Note that, each user may be interested in different features. Therefore, the semantic transmitter needs to send       semantic features according to users' interests,   while the unintended semantic features do not need to be transmitted.

 \subsubsection{Semantic Encoder}

 Based on the  knowledge base which involves shared knowledge between the transmitter and receivers, the semantic encoder extracts  $L$ disentangled semantic    features  $Z = {\left\{ {{z_l}} \right\}_{l \in {\mathcal{L}}}}$  from  the data ${{X_{\rm{d}}}}$, where ${z_l}$ denotes the $l$th semantic    feature, and ${\mathcal{L}} \buildrel \Delta \over = \left\{ {1,...,L} \right\}$. On one hand,  the semantic encoder fulfills the function of a   source encoder, which
compresses the  high-dimensional data ${{X_{\rm{d}}}}$   into low-dimensional semantic  features $Z$.
 On the other hand, the semantic encoder also plays the role of  channel encoder,  which involves redundancy to combat the   channel variations. Note that the  redundancy added in the semantic encoder is to improve the robustness of transmission in the semantic level, while the conventional channel encoder  improves the robustness at the bit level.

 \subsubsection{Features
Selection}

The disentangled semantic features $Z$ could contain  multiple semantic features for multiple users, and users' interested semantic features  may be  different. Therefore, the features selection module selects the   semantic features for each user based on  the  task's  requirement.
  Specifically,
      let    ${\{ {z_l}\} _{l \in {{\mathcal{L}_i}}}}$ denote the    interested features    of User $i$, where  $ {{{\mathcal{L}}_i}}  \subseteq {\mathcal{L}}$.
         Since User $i$  is only  interested   in the    features     ${\{ {z_l}\} _{l \in {{\mathcal{L}_i}}}}$,   the rest of the features $\{ {z_n}\} _{n \in {\mathcal{L}\backslash {{\mathcal{L}_i}}}}$ can be viewed as the ``redundancy" for User $i$.
Then, let ${X_{i,{\rm{s}}}}$ denote the selected semantic features of User $i$, which   is given as
\begin{align}{X_{i,{\rm{s}}}} = {\left\{ {{z_l}} \right\}_{l \in {{\mathcal{L}_i}}}}.
\end{align}
Furthermore,
  the  selected semantic features of $K$ users ${X_{\rm{s}}}$ are given as
\begin{align}{X_{\rm{s}}} = \left\{ {{X_{i,{\rm{s}}}}} \right\}_{i = 1}^K.\end{align}

At last,
 the   selected semantic features ${X_{\rm{s}}}$ are encoded into $X$ through conventional bit-level source coding and channel coding, and then transmitted to the broadcast channel.

 \subsubsection{Semantic Broadcast Channel}

For User $i$,    the received signal $Y_i$ is   decoded by the bit-level channel decoder and the source decoder, and    output  the estimated intended   feature  ${Y_{i,{\rm{s}}}} = {\left\{ {{{\widehat z}_l}} \right\}_{l \in {{\cal L}_{{{i}}}}}}$. The  semantic broadcast channel with input  $X_s$ and outputs $({Y_{1,{\rm{s}}}},\ldots,{Y_{K,{\rm{s}}}})$ can be characterized  as the conditional probability
 \begin{align}
p\left( {{Y_{1,s}}, \ldots ,{Y_{K,s}}|{X_s}} \right)\;{\mkern 1mu}  = &p\left( {X|{X_s}} \right)P\left( {{Y_1}, \ldots ,{Y_K}|X} \right) \prod\limits_{k = 1}^K p \left( {{Y_{k,{\rm{s}}}}|{Y_k}} \right),
\end{align}
 where $P(Y_1,\ldots,Y_K|X)$ is the transitional probability of the broadcast channel with  $X$ being the  channel input  and $Y_k$ being the received signal at receiver $k$.

 \subsubsection{Feature Completion}
Due to the feature selection,  the unintended features are not transmitted, and
  the users only received the interested features.  Consider the case that users aim to retrieve the source data based on the interested features. Then, a feature completion module can be used to help the users obtain the unintended features. Specifically,
let  ${\left\{ {{{\overline z}_t}} \right\}_{l \in {\cal L}\backslash {{\cal L}_i}}}$ denote the unintended  features obtained from the knowledge base.  Based on ${\left\{ {{{\overline z}_t}} \right\}_{l \in {\cal L}\backslash {{\cal L}_i}}}$ and the estimated interested features ${\left\{ {{{\hat z}_l}} \right\}_{l \in {{\cal L}_i}}} $, User $i$ obtains the completed   semantic features   ${{\hat Z}_i} = {\left\{ {{{\hat z}_l}} \right\}_{l \in {{\cal L}_i}}} \cup {\left\{ {{{\overline z }_t}} \right\}_{t \in {\cal L}\backslash {{\cal L}_i}}}$.

        For example,
considering a semantic BC system for   staff clothing images transmission.  The  intended semantic features of  each user may be different, i.e., some users may be interested in  the staff's  hat features, while   some users may be interested in  the staff's  clothing features. There may be  also some uninterested semantic features  such as   staff's gender, skin color, and hairstyle, and background of the image.
  Therefore, the users can generate unintended semantic features based on the shared knowledge base,
  such as the staff's gender, skin color and hairstyle. Note that, the  generated unintended
   semantic features at the receiver may be different from the corresponding features of the  image at the transmitter.
 Then, the user combines the received clothing features with its own generated unintended features.  	
	Moreover, both the simulation and prototype test verification of feature selection and completion  will be presented   in Table V.


 \subsubsection{Semantic Decoder and Semantic Destination}
 With the completed   semantic features   $\widehat Z_i$,
the semantic decoder of User $i$   recovers the data  ${\widehat X_{i,{\rm{d}}}}$ based on the   knowledge base,  and finally sends it to the semantic destination.



So far, the key modules of the proposed features-disentangled broadcast semantic
BC framework have been introduced, and we will further present  a feasible realization and  prototype verification of  this framework in Sections V and VI, respectively.


   \section{Information-Theoretic Metrics of Semantic BCs }


 In this section, we further investigate two information-theoretic metrics for semantic BC: the ultimate compression rate under both the distortion
and perception constraints \cite{Blau_ICML_2019}, and the achievable rate region of the semantic BC. Specifically,  we
  propose the optimal distortion allocation scheme for multi-source data     semantic compression and derive the  achievable rate region for the linear semantic BCs, where    the conditional probability $p\left( {{Y_{1,{\rm{s}}}}, \ldots ,{Y_{K,{\rm{s}}}}|{X_{\rm{s}}}} \right)$ satisfies the following relation
\begin{align}{Y_{i,{\rm{s}}}} = {G_{i}}{X_{\rm{s}}} + {N_{i,{\rm{s}}}}, ~i = 1,...,K,
 \end{align}
 where   ${G_i}$ denotes the \emph{effective  channel gain} of User $i$ from the feature  selection  module to the feature completion module, and
 ${N_{i,{\rm{s}}}}$ denotes the received   semantic  noise.   Since  the physical channel   noise   generally follows Gaussian distribution and  the  source encoder and channel encoder   are  non-linear mapping functions,     the received  semantic    noise  ${N_{i,{\rm{s}}}}$ is assumed to follow a  non-Gaussian distribution with   variance $\sigma _{i,{\rm{s}}}^2$.

   \subsection{ Distortion Allocation for Multi-Source  Data   Semantic Compression}

Most of the existing semantic communications  focused on  a single source  data   semantic compression.
However, in  multi-user semantic BC networks, the interested
  data (or features) of each user may be different, and thus multi-source data  compression is the general case in the multi-user semantic communication networks. The question naturally arises as to how
we should allot this distortion to the  multi-source  data  to minimize the
total distortion under both the distortion and perception constraints.

Hence,  we   investigate the optimal distortion allocation scheme with  ${L_{\rm{d}}}>1$   independent
 sources data   ${\left\{ {{X_{i{\rm{,d}}}}} \right\}_{i = 1}^{L_{\rm{d}}}}$
       for  semantic BC networks.
We aim to optimize distortion allocation for   multi-source  data  compression with both
  squared error distortion and the perception  constraints. Mathematically, the   distortion allocation  optimization  can be formulated as
 \begin{subequations}\label{R_D_P}
\begin{align}
R\left( {D,P} \right) = \mathop {\min }\limits_{\left\{ {p\left( {{{\widehat X}_{i{\rm{,d}}}}|{X_{i{\rm{,d}}}}} \right)} \right\}}~&{\rm{I}}\left( {\left\{ {{X_{i{\rm{,d}}}}} \right\}_{i = 1}^{L_{\rm{d}}};\left\{ {{{\widehat X}_{i{\rm{,d}}}}} \right\}_{i = 1}^{L_{\rm{d}}}} \right)\\
 {\rm{s.t.}}~&\sum\limits_{i = 1}^{L_{\rm{d}}} {{\rm{E}}\left\{ {{{\left\| {{X_{i{\rm{,d}}}} - {{\widehat X}_{i{\rm{,d}}}}} \right\|}^2}} \right\}}  \le D,\\
 &\sum\limits_{i = 1}^{L_{\rm{d}}} {{d_{{\rm{KL}}}^{\left( i \right)}}\left( {{p_{{{\hat X}_{i,{\rm{d}}}}}},{p_{{X_{i,{\rm{d}}}}}}} \right)}  \le P,
\end{align}
\end{subequations}
where  $D$ and $P$ denote the total distortion and   {Kullback-Leibler (KL) divergence thresholds, respectively.

In this paper, we consider  the multiple independent Gaussian distributed data   ${\left\{ {{X_{i{\rm{,d}}}}} \right\}_{i = 1}^{L_{\rm{d}}}}$,
       i.e., ${X_{i,{\rm{d}}}} \sim  \mathcal{N}\left( {0,\sigma _{i,{\rm{d}}}^2} \right)$.
   Moreover, let ${{D}_i}$ denote the squared-error   distortion between ${X_{i,{\rm{d}}}}$ and ${{\hat X}_{i,{\rm{d}}}}$, i.e.,
     \begin{align}{{D}_i} = {\rm{E}}\left\{ {{{\left\| {{X_{i,{\rm{d}}}} - {{\hat X}_{i,{\rm{d}}}}} \right\|}^2}} \right\}.
     \end{align}

  Then, the mutual information ${\rm{I}}\left( {\left\{ {{X_{i{\rm{,d}}}}} \right\}_{i = 1}^{L_{\rm{d}}};\left\{ {{{\widehat X}_{i{\rm{,d}}}}} \right\}_{i = 1}^{L_{\rm{d}}}} \right)$ can be written as  \cite{Cover_Book}
 \begin{align}{\rm{I}}\left( {\left\{ {{X_{i,{\rm{d}}}}} \right\}_{i = 1}^{{L_{\rm{d}}}};\left\{ {{{\hat X}_{i,{\rm{d}}}}} \right\}_{i = 1}^{{L_{\rm{d}}}}} \right) = \sum\limits_{i = 1}^{{L_{\rm{d}}}} {{{\left[ {\frac{1}{2}\log \left( {\frac{{\sigma _{i,{\rm{d}}}^2}}{{{D_i}}}} \right)} \right]}^ + }},
\end{align}
where ${\widehat X_{i,{\rm{d}}}}$ follows a Gaussian distribution, i.e.,   ${\widehat X_{i,{\rm{d}}}} \sim \mathcal{N}\left( {0,\sigma _{i{\rm{,d}}}^2 - {D_i}} \right)$, and    ${\left[ x \right]^ + } = x$ if $x \ge 0$, otherwise, ${\left[ x \right]^ + } = 0$.

Thus, for the Gaussian distributed data ${X_{i,{\rm{d}}}}$ and the reconstructed data ${\widehat X_{i,{\rm{d}}}}$,
  the KL-divergence ${d_{{\rm{KL}}}^{\left( i \right)}}\left( {{p_{{X_{i{\rm{,d}}}}}},{p_{{{\widehat X}_{i{\rm{,d}}}}}}} \right)$  is given as
\begin{align}{d_{{\rm{KL}}}^{\left( i \right)}}\left( {{p_{{{\hat X}_{i,{\rm{d}}}}}},{p_{{X_{i,{\rm{d}}}}}}} \right) = \frac{1}{2}\left( {\ln \frac{{\sigma _{i,{\rm{d}}}^2}}{{\sigma _{i,{\rm{d}}}^2 - {D_i}}} + \frac{{\sigma _{i,{\rm{d}}}^2 - {D_i}}}{{\sigma _{i,{\rm{d}}}^2}} - 1} \right).\end{align}


Thus, the optimal distortion allocation    problem \eqref{R_D_P} can be reformulated as
\begin{subequations}\label{RDP2}
\begin{align}
R\left( {D,P} \right) = \mathop {\min }\limits_{\left\{ {{D_i}} \right\}_{i = 1}^{{L_{\rm{d}}}}} &\sum\limits_{i = 1}^{{L_{\rm{d}}}} {{{\left[ {\frac{1}{2}\log \left( {\frac{{\sigma _{i,{\rm{d}}}^2}}{{{D_i}}}} \right)} \right]}^ + }}  \\
{\rm{s}}.{\rm{t}}.&\sum\limits_{i = 1}^{L_{\rm{d}}} {{D_i}}  \le D,\label{RDP2b}\\
&\sum\limits_{i = 1}^{L_{\rm{d}}} {\frac{1}{2}\left( {\ln \frac{{\sigma _{i,{\rm{d}}}^2}}{{\sigma _{i,{\rm{d}}}^2 - {D_i}}} + \frac{{\sigma _{i,{\rm{d}}}^2 - {D_i}}}{{\sigma _{i,{\rm{d}}}^2}} - 1} \right)}  \le P.\label{RDP2c}
\end{align}
\end{subequations}

Note that, problem \eqref{RDP2} is convex in $\left\{ {{{{D}}_i}} \right\}$, and the Lagrangian function of problem \eqref{RDP2} $L\left( {\left\{ {{D_i}} \right\},{\lambda _D},{\lambda _P}} \right)$ is given as
\begin{align}
L\left( {\left\{ {{D_i}} \right\},{\lambda _D},{\lambda _P}} \right) =& \sum\limits_{i = 1}^{L_{\rm{d}}} {\frac{1}{2}\ln \frac{{{\sigma _{i,{\rm{d}}}^2}}}{{{D_i}}} + {\lambda _D}} \left( {\sum\limits_{i = 1}^{L_{\rm{d}}} {{D_i}}  - D} \right)\nonumber\\
 &+ {\lambda _P}\left( {\frac{1}{2}\sum\limits_{i = 1}^{L_{\rm{d}}} {\left( {\ln \frac{{{\sigma _{i,{\rm{d}}}^2} - {D_i}}}{{{\sigma _{i,{\rm{d}}}^2}}} + \frac{{{\sigma _{i,{\rm{d}}}^2}}}{{{\sigma _{i,{\rm{d}}}^2} - {D_i}}} - 1} \right)}  - P} \right),
\end{align}
where ${\lambda _D} \ge 0$ and ${\lambda _P} \ge 0$ are Lagrange multipliers
  associated with   constraints \eqref{RDP2b} and \eqref{RDP2c}, respectively.

Furthermore, let
  the first derivative of the function $L\left( {\left\{ {{D_i}} \right\},{\lambda _D},{\lambda _P}} \right)$ with respect to ${D_i}$ be equal to 0, i.e.,
 \begin{align} { - \frac{1}{{2{D_i}}} + {\lambda _D} + {\lambda _P}\frac{1}{2}\frac{{{D_i}}}{{{\sigma _{i,{\rm{d}}}^2}\left( {{\sigma _{i,{\rm{d}}}^2} - {D_i}} \right)}} = 0,~i = 1,..,{L_{\rm{d}}}}.
 \end{align}
Thus, the optimal distortions  ${D_i}$ are given as
\begin{align}{D_i} = \frac{{{\sigma _{i,{\rm{d}}}^2}\left( {\sqrt {{{\left( {1 - 2{\lambda _D}{\sigma _{i,{\rm{d}}}^2}} \right)}^2} + 4{\lambda _P}}  - \left( {1 + 2{\lambda _D}{\sigma _{i,{\rm{d}}}^2}} \right)} \right)}}{{2\left( {{\lambda _P} - 2{\lambda _D}{\sigma _{i,{\rm{d}}}^2}} \right)}},
\end{align}
where  the parameters $\lambda _D$ and $\lambda _P$ are the solutions of the following equations
\begin{subequations}
\begin{align}
&{{\lambda _D}\left( {\sum\limits_{i = 1}^K {{D_i}}  - D} \right) = 0},\\
&{{\lambda _P}\left( {\frac{1}{2}\sum\limits_{i = 1}^{L_{\rm{d}}} {\left( {\ln \frac{{{\sigma _{i,{\rm{d}}}^2} - {D_i}}}{{{\sigma _{i,{\rm{d}}}^2}}} + \frac{{{\sigma _{i,{\rm{d}}}^2}}}{{{\sigma _{i,{\rm{d}}}^2} - {D_i}}} - 1} \right)}  - P} \right) = 0.}
\end{align}
\end{subequations}

\begin{figure}
	\centering
	\includegraphics[width=0.6\textwidth]{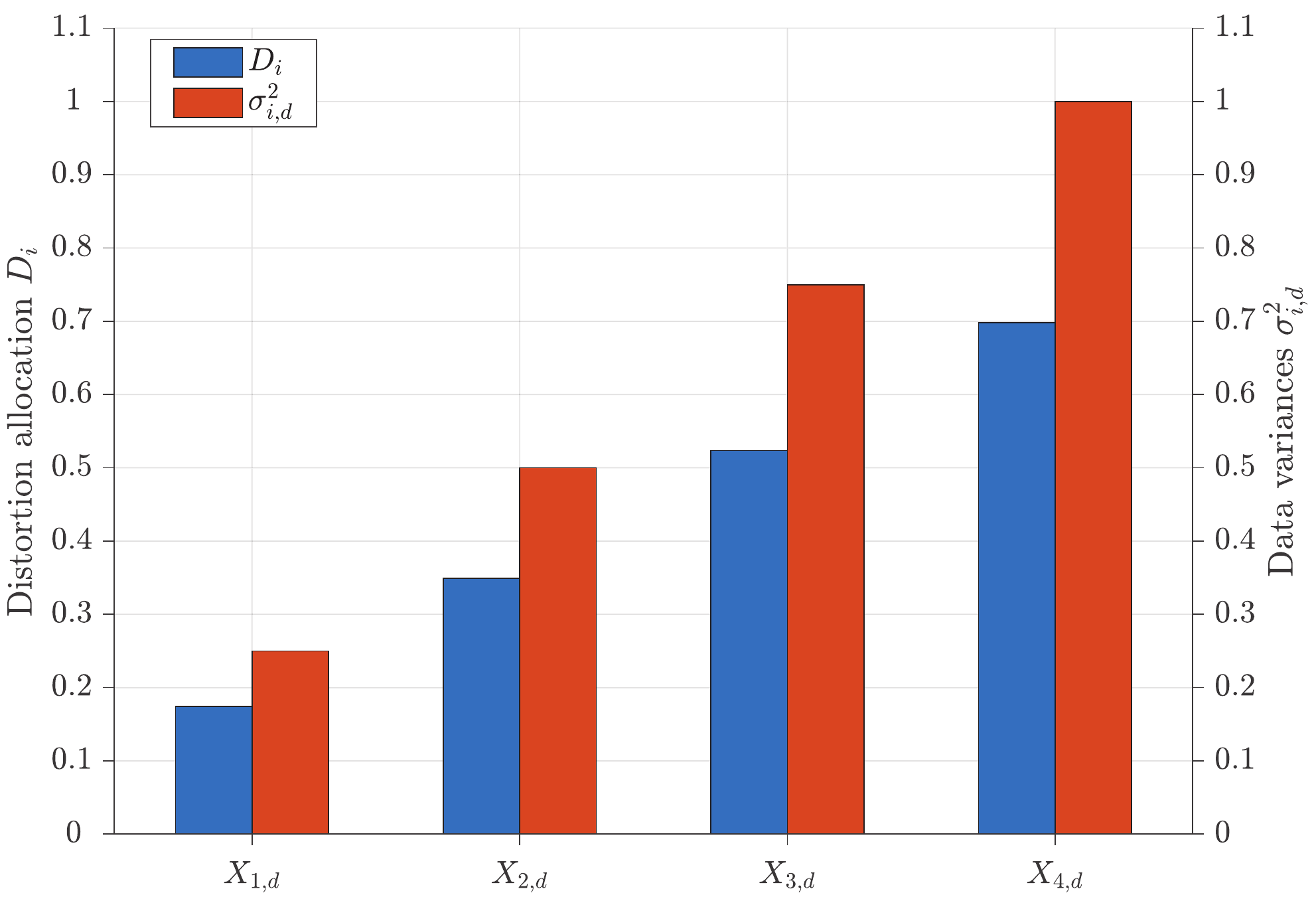}
	\caption{Distortion allocation for Gaussian distributed data  $\left\{ {{{\rm{X}}_{i,{\rm{d}}}}} \right\}_{i = 1}^4$   with   variances $\sigma _{1,{\rm{d}}}^2 = 0.25$, $\sigma _{i,{\rm{d}}}^2 = 0.5$, $\sigma _{i,{\rm{d}}}^2 = 0.75$,and $\sigma _{i,{\rm{d}}}^2 = 1$, respectively.}
	\label{Distortion_Allo}
\end{figure}

Thus, the   rate distortion function of the data ${X_{i,{\rm{d}}}}$  is given as
\begin{align}{R_i} = {\left[ {\frac{1}{2}\ln \frac{{2\left( {{\lambda _P} - 2{\lambda _D}\sigma _{i,{\rm{d}}}^2} \right)}}{{\sqrt {{{\left( {1 - 2{\lambda _D}\sigma _{i,{\rm{d}}}^2} \right)}^2} + 4{\lambda _P}}  - \left( {1 + 2{\lambda _D}\sigma _{i,{\rm{d}}}^2} \right)}}} \right]^ + },
\end{align}where $i = 1,..,{L_{\rm{d}}}$. Fig. \ref{Distortion_Allo} shows the optimal distortion allocation for Gaussian distributed data  $\left\{ {{{\rm{X}}_{i,{\rm{d}}}}} \right\}_{i = 1}^4$   with   variances $\sigma _{1,{\rm{d}}}^2 = 0.25$, $\sigma _{i,{\rm{d}}}^2 = 0.5$, $\sigma _{i,{\rm{d}}}^2 = 0.75$, and $\sigma _{i,{\rm{d}}}^2 = 1$, respectively. From Fig. \ref{Distortion_Allo},  we   observe that with a larger data variance  $\sigma _{i,{\rm{d}}}^2$, e.g.,
N = 50, the   allocated distortion   is more, i.e., the  allocated distortion increases with the values of variance of the data. 

   \subsection{Achievable Rate Region   of  2-Receiver Semantic Broadcast Channel}

Due to the   non-Gaussian
distributed semantic channel noise, the classical Shannon capacity region results of the broadcast channel   (based on
Gaussian distributed noise) cannot directly be applied in  semantic broadcast channels.

In the following,  we establish an achievable rate region  of the  2-user degraded semantic BCs with $|{G_1}| \geq |{G_2}|$ and $\sigma_{1,s}=\sigma_{2,s}$, which implies that User 2 observes stronger signals than that of User 1.
   The main idea is that the broadcast
transmitter employs semantic features splitting and  superposition, and users  employ
 successive  interference  cancelation (SIC)\cite{Cover_Book}.
   Specifically, since ${G_1} > {G_2}$,
  the semantic feature ${X_{{\rm{2}},{\rm{s}}}}$ for user $2$, which  can be viewed as common message, is encoded into the
cloud center signal, while  the semantic feature ${X_{{\rm{1}},{\rm{s}}}}$, which  can be viewed as     private message, is encoded into the satellite signal.
 Let ${R_{1,{\rm{s}}}}$ and  ${R_{2,{\rm{s}}}}$ denote the information rates from the
feature selection module of the transmitter to the feature completion module of Users $1$ and $2$, respectively.
A rate pair $\left( {{{R_{1,{\rm{s}}}}},{{R_{2,{\rm{s}}}}}} \right)$ is achievable for the 2-user degraded semantic BCs if it satisfies the conditions \cite{Cover_Book}
\begin{subequations}
\begin{align}
&{R_{1,{\rm{s}}}} \le {\rm{I}}\left( {{X_{{\rm{1,s}}}};{Y_{1{\rm{,s}}}}|{X_{{\rm{2,s}}}}} \right),\\
&{R_{2,{\rm{s}}}} \le {\rm{I}}\left( {{X_{{\rm{2,s}}}};{Y_{2{\rm{,s}}}}} \right).
\end{align}
\end{subequations}

Moreover, let $P$ denote  the total transmission  power of the transmitter, and
${\alpha}P$ and ${\left( {1 - \alpha } \right)P}$ denote the allocated power   to the semantic feature  ${X_{{\rm{1}},{\rm{s}}}}$  and
  ${\left( {1 - \alpha } \right)P}$,  respectively,
  where  ${\alpha} \in \left[ {0,1} \right]$ is a power allocation factor.
Thus, the transmitted signal $X$ of the semantic broadcast network is given as
 \begin{align}
X = \sqrt {\alpha P} {X_{{\rm{1}},{\rm{s}}}} + \sqrt {\left( {1 - \alpha } \right)P} {X_{{\rm{2}},{\rm{s}}}}.\end{align}

The received signals at User $1$ and User $2$ are, respectively, given as
\begin{subequations}
\begin{align}
{Y_{{\rm{1}},{\rm{s}}}} = {G_1}\sqrt {\alpha P} {X_{{\rm{1}},{\rm{s}}}} + {G_1}\sqrt {\left( {1 - \alpha } \right)P} {X_{{\rm{2}},{\rm{s}}}} + {N_{{\rm{1}},{\rm{s}}}},\\
{Y_{2,{\rm{s}}}} = {G_2}\sqrt {\alpha P} {X_{{\rm{1}},{\rm{s}}}} + {G_2}\sqrt {\left( {1 - \alpha } \right)P} {X_{{\rm{2}},{\rm{s}}}} + {N_{{\rm{2}},{\rm{s}}}}.
\end{align}
\end{subequations}

User $1$ utilizes  SIC     to
   decode the semantic feature ${X_{{\rm{2}},{\rm{s}}}}$  first and cancel ${X_{{\rm{2}},{\rm{s}}}}$  from the received signal. Then, User $1$ decodes the      semantic feature    ${X_{{\rm{1}},{\rm{s}}}}$.  After SIC,  the residual received signal of User $1$
  is given  as
\begin{align}Y_{{\rm{1}},{\rm{s}}}^{{\rm{SIC}}} = {G_1}\sqrt {{\alpha }P} {X_{{\rm{1}},{\rm{s}}}} + {N_{{\rm{1}},{\rm{s}}}}\label{Received_sic}
\end{align}

While User $2$ can only decode the semantic feature ${X_{{\rm{2}},{\rm{s}}}}$, and the      semantic feature  {${X_{{\rm{1}},{\rm{s}}}}$ is the interference for user $2$.

\begin{Lemma}[Achievable rate region of degraded semantic broadcast channels]
Consider the degraded semantic broadcast channel, where ${G_1} > {G_2}$,  the
achievable rates  ${R_{1,{\rm{s}}}}$ and  ${R_{2,{\rm{s}}}}$ are bounded by
\begin{subequations}\label{rate_region}
\begin{align}
&R_{{\rm{1}},{\rm{s}}}^{{\rm{eq}}} \le {R_{1,{\rm{s}}}} \le R_{{\rm{1}},{\rm{s}}}^{{\rm{eq}}} + {d_{{\rm{KL}}}}\left( {{p_{{N_{{\rm{1}},{\rm{s}}}}}}\left( x \right),{p_{N_{{\rm{1}},{\rm{s}}}^{{\rm{eq}}}}}\left( x \right)} \right),\\
&R_{2,{\rm{s}}}^{{\rm{eq}}} \le {R_{2,{\rm{s}}}} \le R_{2,{\rm{s}}}^{{\rm{eq}}} + {d_{{\rm{KL}}}}\left( {{p_{{N_{2,{\rm{s}}}}}}\left( x \right),{p_{N_{2,{\rm{s}}}^{{\rm{eq}}}}}\left( x \right)} \right),
 \end{align}
 \end{subequations}
 where  $N_{1,{\rm{s}}}^{{\rm{eq}}}$ and $N_{2,{\rm{s}}}^{{\rm{eq}}}$ are the equivalent Gaussian distributed   noises with the same variances as ${N_{{\rm{1}},{\rm{s}}}}$ and ${N_{{\rm{1}},{\rm{s}}}}$, respectively, i.e.,  ${\mathop{\rm var}} \left( {N_{1,{\rm{s}}}^{{\rm{eq}}}} \right) = \sigma _{1,{\rm{s}}}^2$, and ${\mathop{\rm var}} \left( {N_{2,{\rm{s}}}^{{\rm{eq}}}} \right) = \sigma _{2,{\rm{s}}}^2$,
 ${d_{{\rm{KL}}}}\left( {p\left( x \right),q\left( x \right)} \right) = \int_{ - \infty }^\infty  {p\left( x \right)\log \frac{{p\left( x \right)}}{{q\left( x \right)}}} {\rm{d}}x$, and
   \begin{subequations}
  \begin{align}
&R_{{\rm{1}},{\rm{s}}}^{{\rm{eq}}} = \frac{1}{2}{\log _2}\left( {1 + \frac{{G_1^2{\alpha _1}P}}{{\sigma _{1,s}^2}}} \right),\\
&R_{{\rm{2}},{\rm{s}}}^{{\rm{eq}}} = \frac{1}{2}{\log _2}\left( {1 + \frac{{G_2^2\left( {1 - \alpha } \right)P}}{{G_2^2{\alpha _1}P + \sigma _{2,s}^2}}} \right).
 \end{align}
\end{subequations}
\end{Lemma}
		\emph{Proof:}
We first introduce  the equivalent Gaussian distributed channel noises $N_{1,{\rm{s}}}^{{\rm{eq}}}$ and $N_{2,{\rm{s}}}^{{\rm{eq}}}$ with the same variances as ${N_{{\rm{1}},{\rm{s}}}}$ and ${N_{{{2}},{\rm{s}}}}$, respectively, i.e.,  ${\mathop{\rm var}} \left( {N_{1,{\rm{s}}}^{{\rm{eq}}}} \right) = \sigma _{1,{\rm{s}}}^2$, and ${\mathop{\rm var}} \left( {N_{2,{\rm{s}}}^{{\rm{eq}}}} \right) = \sigma _{2,{\rm{s}}}^2$. The PDFs of  $N_{1,{\rm{s}}}^{{\rm{eq}}}$ and $N_{2,{\rm{s}}}^{{\rm{eq}}}$  are, respectively, given as
 \begin{subequations}
  \begin{align}
{p_{N_{1,{\rm{s}}}^{{\rm{eq}}}}}\left( x \right) = \frac{1}{{\sqrt \pi  {\sigma _{1,s}}}}\exp \left( { - \frac{{{x^2}}}{{\sigma _{1,s}^2}}} \right),\\
{p_{N_{2,{\rm{s}}}^{{\rm{eq}}}}}\left( x \right) = \frac{1}{{\sqrt \pi  {\sigma _{2,s}}}}\exp \left( { - \frac{{{x^2}}}{{\sigma _{2,s}^2}}} \right).
\end{align}
\end{subequations}

With the equivalent Gaussian distributed channel noise $N_{1,{\rm{s}}}^{{\rm{eq}}}$ and $N_{2,{\rm{s}}}^{{\rm{eq}}}$, the corresponding rates of User $1$ and User $2$   are, respectively, given as
  \begin{subequations}
  \begin{align}
&R_{{\rm{1}},{\rm{s}}}^{{\rm{eq}}} = \frac{1}{2}{\log _2}\left( {1 + \frac{{G_1^2{\alpha _1}P}}{{\sigma _{1,s}^2}}} \right),\\
&R_{{\rm{2}},{\rm{s}}}^{{\rm{eq}}} = \frac{1}{2}{\log _2}\left( {1 + \frac{{G_2^2\left( {1 - \alpha } \right)P}}{{G_2^2{\alpha _1}P + \sigma _{2,s}^2}}} \right).
 \end{align}
\end{subequations}

For the non-Gaussian distributed channel noise ${N_{{{1}},{\rm{s}}}}$ and ${N_{{{2}},{\rm{s}}}}$, the
semantic communication  rates  $R_1$ and $R_2$ are bounded by \cite{Ihara}
\begin{subequations}
\begin{align}
&R_{{\rm{1}},{\rm{s}}}^{{\rm{eq}}} \le {R_{1,{\rm{s}}}} \le R_{{{1}},{\rm{s}}}^{{\rm{eq}}} + {d_{{\rm{KL}}}}\left( {{p_{{N_{{{1}},{\rm{s}}}}}}\left( x \right),{p_{N_{{{1}},{\rm{s}}}^{{\rm{eq}}}}}\left( x \right)} \right),\\
&R_{2,{\rm{s}}}^{{\rm{eq}}} \le {R_{2,{\rm{s}}}} \le R_{2,{\rm{s}}}^{{\rm{eq}}} + {d_{{\rm{KL}}}}\left( {{p_{{N_{2,{\rm{s}}}}}}\left( x \right),{p_{N_{2,{\rm{s}}}^{{\rm{eq}}}}}\left( x \right)} \right).
 \end{align}
 \end{subequations}
\qed

\begin{figure}[htbp]
    \centering
    \begin{minipage}[t]{0.5\textwidth}
        \centering
        \includegraphics[width=\textwidth]{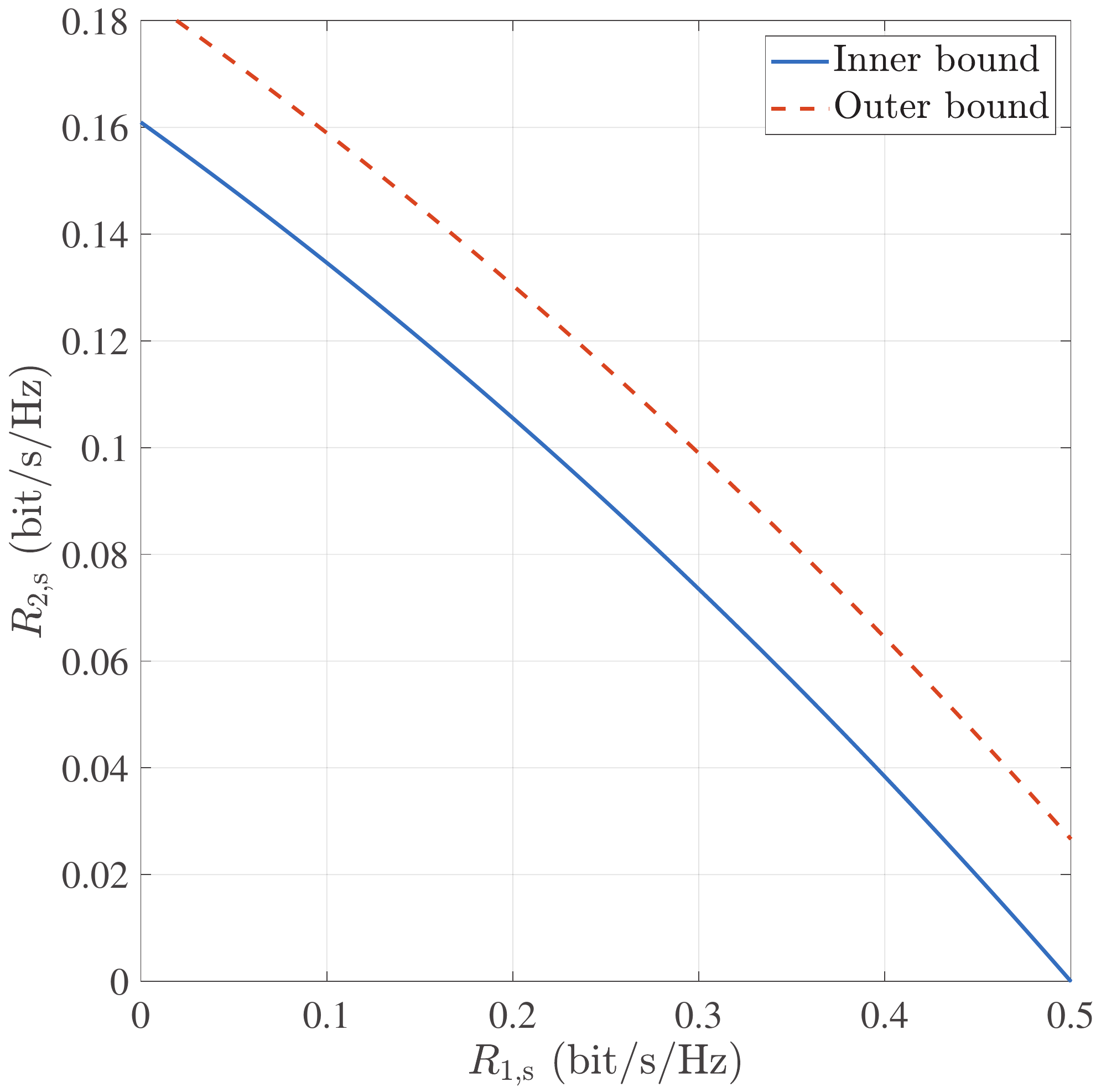}
        \vskip-0.2cm\centering {\footnotesize (a)}
    \end{minipage}
    \begin{minipage}[t]{0.5\textwidth}
        \centering
        \includegraphics[width=\textwidth]{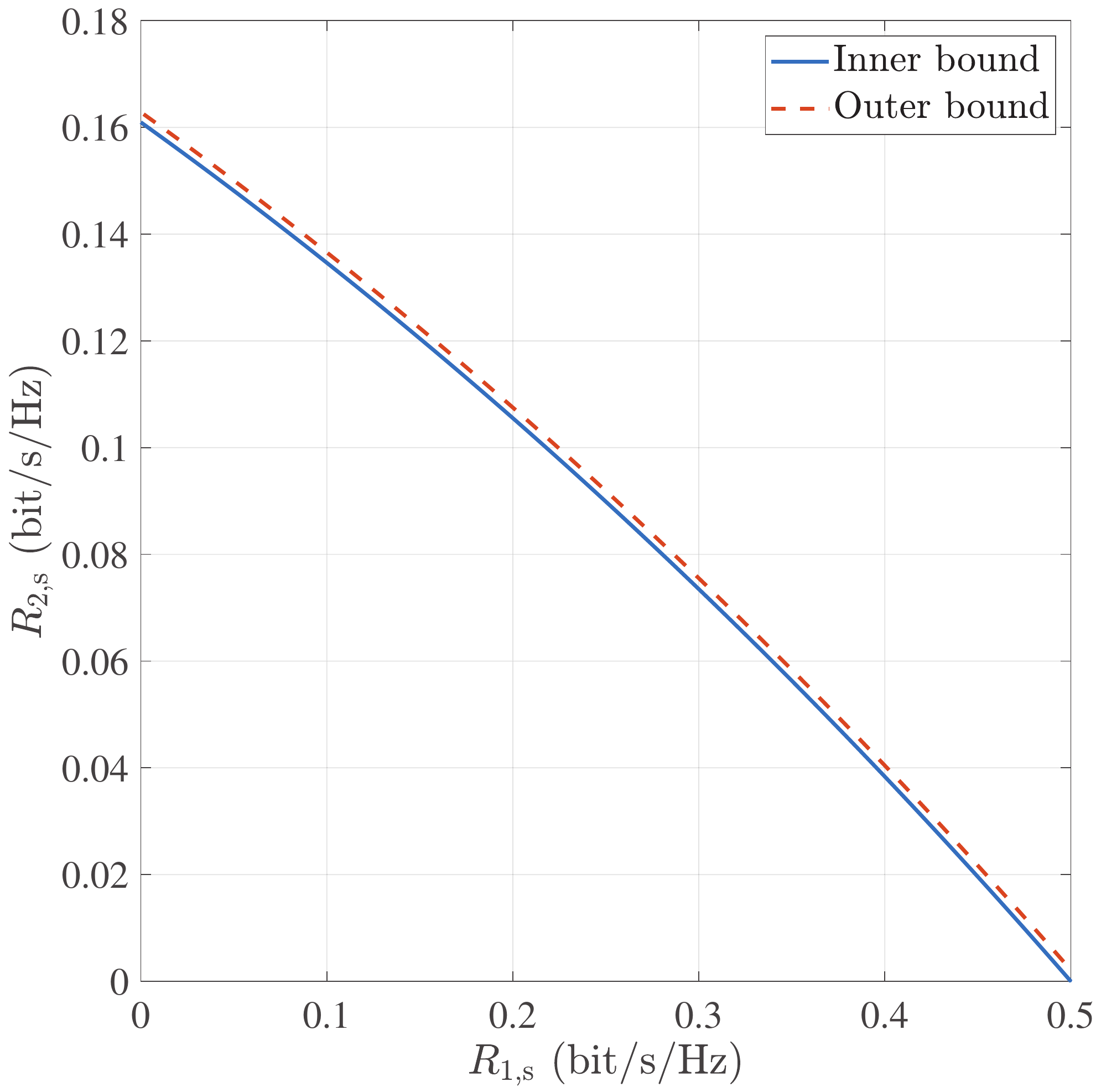}
        \vskip-0.2cm\centering {\footnotesize (b)}
    \end{minipage}
   \caption{
  (a) Inner and outer bounds of semantic broadcast rate region   with ${p_{{N_{{\rm{1}},{\rm{s}}}}}}\left( x \right) = {p_{{N_{2,{\rm{s}}}}}}\left( x \right) = {q_1}\left( x \right)$;
   (b) Inner and outer bounds of semantic broadcast rate region   with ${p_{{N_{{\rm{1}},{\rm{s}}}}}}\left( x \right) = {p_{{N_{2,{\rm{s}}}}}}\left( x \right) = {q_2}\left( x \right)$.}
   \label{capactiy of rate}
\end{figure}

In the following, we   numerically demonstrate the derived achievable rates region results
  \eqref{rate_region} of the semantic broadcast channel with two different PDFs ${q_1}\left( x \right)$ and   ${q_2}\left( x \right)$ cases with the same variances, i.e.,  ${q_1}\left( x \right) \buildrel \Delta \over = \frac{1}{{2\sqrt 3 }}\left( {{\rm{erf}}\left( {\frac{1}{{\sqrt 2 }} - \sqrt {\frac{2}{3}} x} \right) + {\rm{erf}}\left( {\frac{1}{{\sqrt 2 }} + \sqrt {\frac{2}{3}} x} \right)} \right)$
  and
  ${q_2}\left( x \right) \buildrel \Delta \over = \frac{1}{{2\sqrt 3 }}\left( {{\rm{erf}}\left( {\frac{{3\sqrt 2 }}{5} - \sqrt {\frac{2}{3}} x} \right) + {\rm{erf}}\left( {\frac{{3\sqrt 2 }}{5} + \sqrt {\frac{2}{3}} x} \right)} \right)$.
  Fig. \ref{capactiy of rate} (a) and  (b) depict   the  inner and outer bounds of the semantic broadcast rate region   with    ${p_{{N_{{\rm{1}},{\rm{s}}}}}}\left( x \right) = {p_{{N_{2,{\rm{s}}}}}}\left( x \right) = {q_1}\left( x \right)$  and ${p_{{N_{{\rm{1}},{\rm{s}}}}}}\left( x \right) = {p_{{N_{2,{\rm{s}}}}}}\left( x \right) = {q_2}\left( x \right)$ respectively,
 where ${{{G}}_1} = 1$ and  ${{{G}}_2} = 0.5$.
 Comparing   Fig. \ref{capactiy of rate} (a) and  (b), it can be
observed that the gap between the inner bound and outer bound is significantly tighter in Fig. \ref{capactiy of rate}  (b). The reason is that the KL ${d_{{\rm{KL}}}}\left( {{p_{{N_{{\rm{1}},{\rm{s}}}}}}\left( x \right),{p_{N_{{\rm{1}},{\rm{s}}}^{{\rm{eq}}}}}\left( x \right)} \right)$ is smaller than that in Fig. \ref{capactiy of rate} (a) case, i.e., the distributions ${p_{{N_{{\rm{1}},{\rm{s}}}}}}\left( x \right) = {p_{{N_{2,{\rm{s}}}}}}\left( x \right) =  {q_2}\left( x \right)$ are closer to the Gaussian distribution than ${p_{{N_{{\rm{1}},{\rm{s}}}}}}\left( x \right) = {p_{{N_{2,{\rm{s}}}}}}\left( x \right) =  {q_1}\left( x \right)$.
Moreover, when the KL divergence   tends to $0$, i.e., ${d_{{\rm{KL}}}}\left( {{p_{{N_{{\rm{1}},{\rm{s}}}}}}\left( x \right),{p_{N_{{\rm{1}},{\rm{s}}}^{{\rm{eq}}}}}\left( x \right)} \right) \to 0$,   the gap between the inner  bound   and the upper bound in  \eqref{rate_region} tends to 0.

\begin{figure*}
	\centering
	\includegraphics[width=1.0\textwidth]{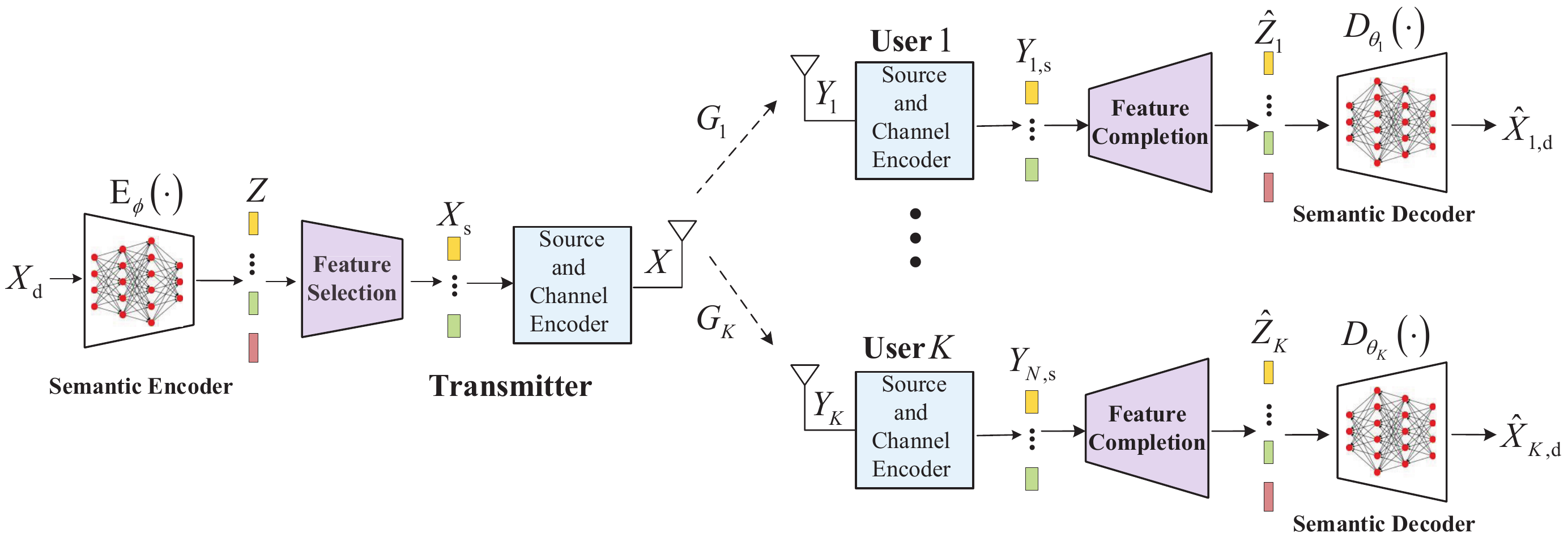}
	\caption{Proposed
		features-disentangled   semantic BC network}
	\label{FD_SM}
\end{figure*}

Note that, although we have derived   two information theoretic
metrics for semantic BC, based on some strong prior information, these derivations are based on some strong prior information, such as the probability distribution of the  semantic information  is assumed known, and the verification of these theories will be explored in the future.

  \section{ Features Disentangled   Semantic BC System Design}
\subsection{Supervised AE Based  Semantic Broadcast  System Design}

Based on the proposed semantic broadcast    framework in Fig. 2,
we propose a features-disentangled  semantic BC network  design,  as shown in Fig. \ref{FD_SM},  which includes a single semantic encoder   ${{\rm{E}}_\phi }\left(  \cdot  \right)$ with    parameters set $\phi$,
   and $K$ semantic decoders $\left\{ {{{\rm{D}}_{{\theta _i}}}\left(  \cdot  \right)} \right\}_{i = 1}^K$  with parameters sets  $\left\{ {{\theta _i}} \right\}_{i = 1}^K$.
 More specifically,     the transmitter with the   encoder network ${{\rm{E}}_\phi }\left(  \cdot  \right)$  extracts the semantic features from the source  data and  disentangles them into multiple independent  and   interpretable
  semantic features.  Then, by applying semantic features selection, the transmitter broadcasts the semantic features   to the intended  users.
  After receiving the intended semantic features,     User $i$    completes the unintended  features with the help of the knowledge base, and
    decodes the   semantic features  with the  decoder network ${{\rm{D}}_{{\theta _i}}}\left(  \cdot  \right)$, where $i = 1,...,K$.

Specifically, to achieve  controllable disentangled semantic features learning,
we exploit the  supervised AE to train the semantic broadcast network in  three steps:
    self reconstruction, exchange common features,   and  exchange different features, which is
  motivated by the
group-supervised learning strategy \cite{Ge_ICLR_2021}. In the following, we will introduce the three training steps in details.

\subsubsection{Self Reconstruction}

 As shown in Fig. \ref{SR} (a), the  self reconstruction training can be regarded as a regular term to ensure that all the semantic information of the input data    can be encoded into   latent semantic features     to avoid information loss.
Specifically, the semantic encoder ${{\rm{E}}_\phi }$ compresses the  $l$th   data sample  ${X_{\rm{d}}^{\left( l \right)}}$ into a latent semantic feature   vector   ${{\rm{Z}}^{\left( l \right)}}$, i.e.,
\begin{align}{{\rm{Z}}^{\left( l \right)}} = {{\rm{E}}_\phi }\left( {X_{\rm{d}}^{\left( l \right)}} \right),
\end{align}
where $l=1,...,{L_{\rm{s}}}$, and ${L_{\rm{s}}}$ denotes the total amount of the sampled data.

Then, the semantic feature vector ${{\rm{Z}}^{\left( l \right)}}$ is broadcasted to $K$ semantic users.
  At the user end,    the  received semantic features  of  User $i$ are given as
\begin{align}\widehat {\rm{Z}}_i^{\left( l \right)} = {G_i}{{\rm{Z}}^{\left( l \right)}} + {N_{i,{\rm{s}}}},~i = 1,...,K.
\end{align}
Then,  User $i$  decodes the received semantic features $\widehat {\rm{Z}}_i^{\left( l \right)}$ and obtains the reconstructed data $\widehat X_{i{\rm{,d}}}^{\left( l \right)} $ as follows
 \begin{align}\widehat X_{i{\rm{,d}}}^{\left( l \right)} = {{\rm{D}}_{{\theta _i}}}\left( {\widehat {\rm{Z}}_i^{\left( l \right)}} \right),~i = 1,...,K.\end{align}

Finally, the semantic encoder   ${{\rm{E}}_\phi }\left(  \cdot  \right)$ with    parameters set $\phi$,
and $K$ semantic decoders
$\left\{ {{{\rm{D}}_{{\theta _i}}}\left(  \cdot  \right)} \right\}_{i = 1}^K$  with parameters sets  $\left\{ {{\theta _i}} \right\}_{i = 1}^K$ are jointly optimized based  on the
standard AE reconstruction loss ${L_{{\rm{SAE,1}}}}$  as follows
\begin{align}{L_{{\rm{AE}},{\rm{1}}}} = \sum\limits_{l = 1}^{{L_{\rm{s}}}} {\sum\limits_{i = 1}^K {{{\left\| {X_{\rm{d}}^{\left( l \right)} - \hat X_{i,{\rm{d}}}^{\left( l \right)}} \right\|}^2}} } .  \end{align}

\subsubsection{Common Semantic Features Exchange}

\begin{figure}[htbp]
    \centering
    \begin{minipage}[t]{0.55\textwidth}
        \centering
        \includegraphics[width=\textwidth]{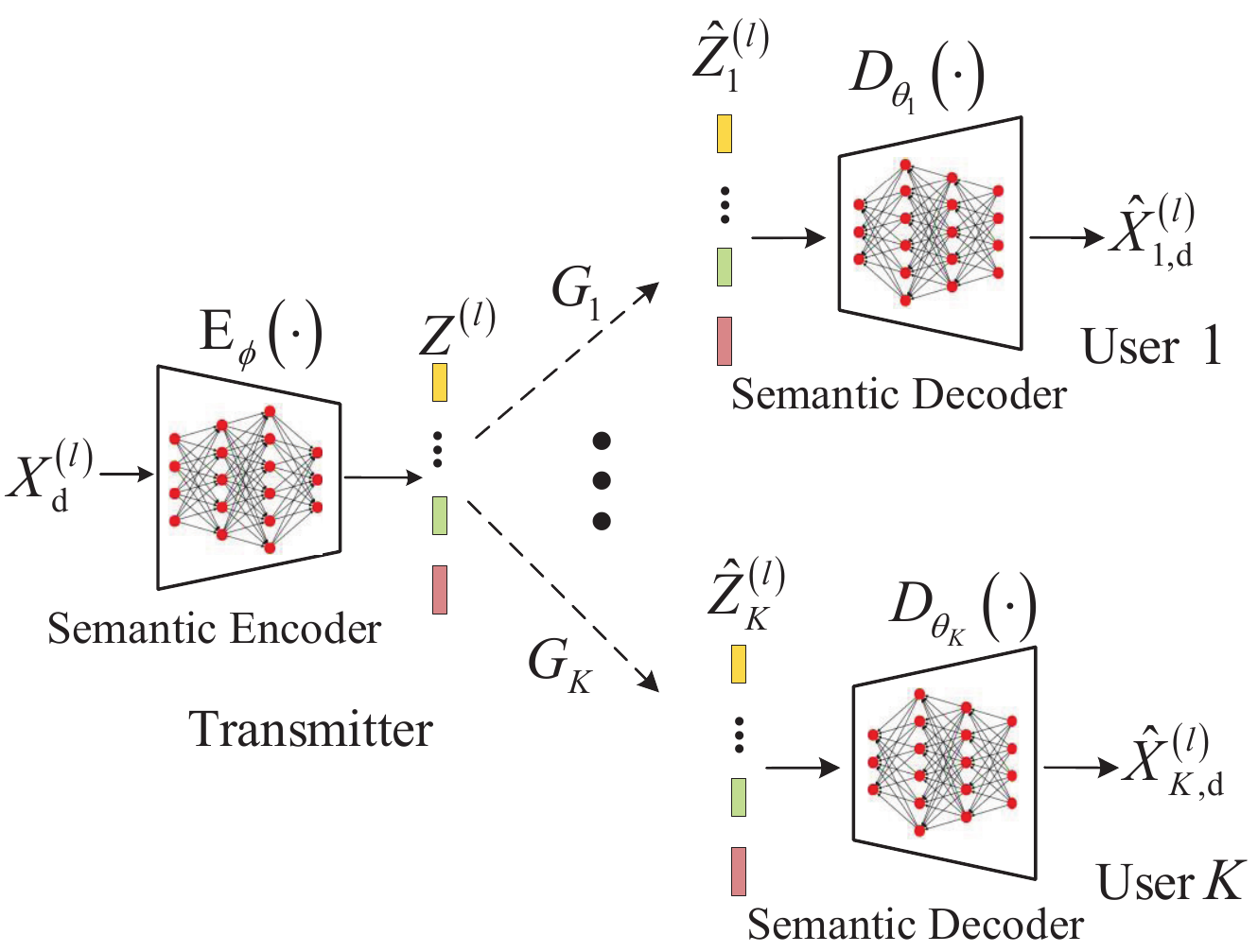}
        \vskip0.2cm\centering {\footnotesize (a)}
    \end{minipage}
    \begin{minipage}[t]{0.55\textwidth}
        \centering
        \includegraphics[width=\textwidth]{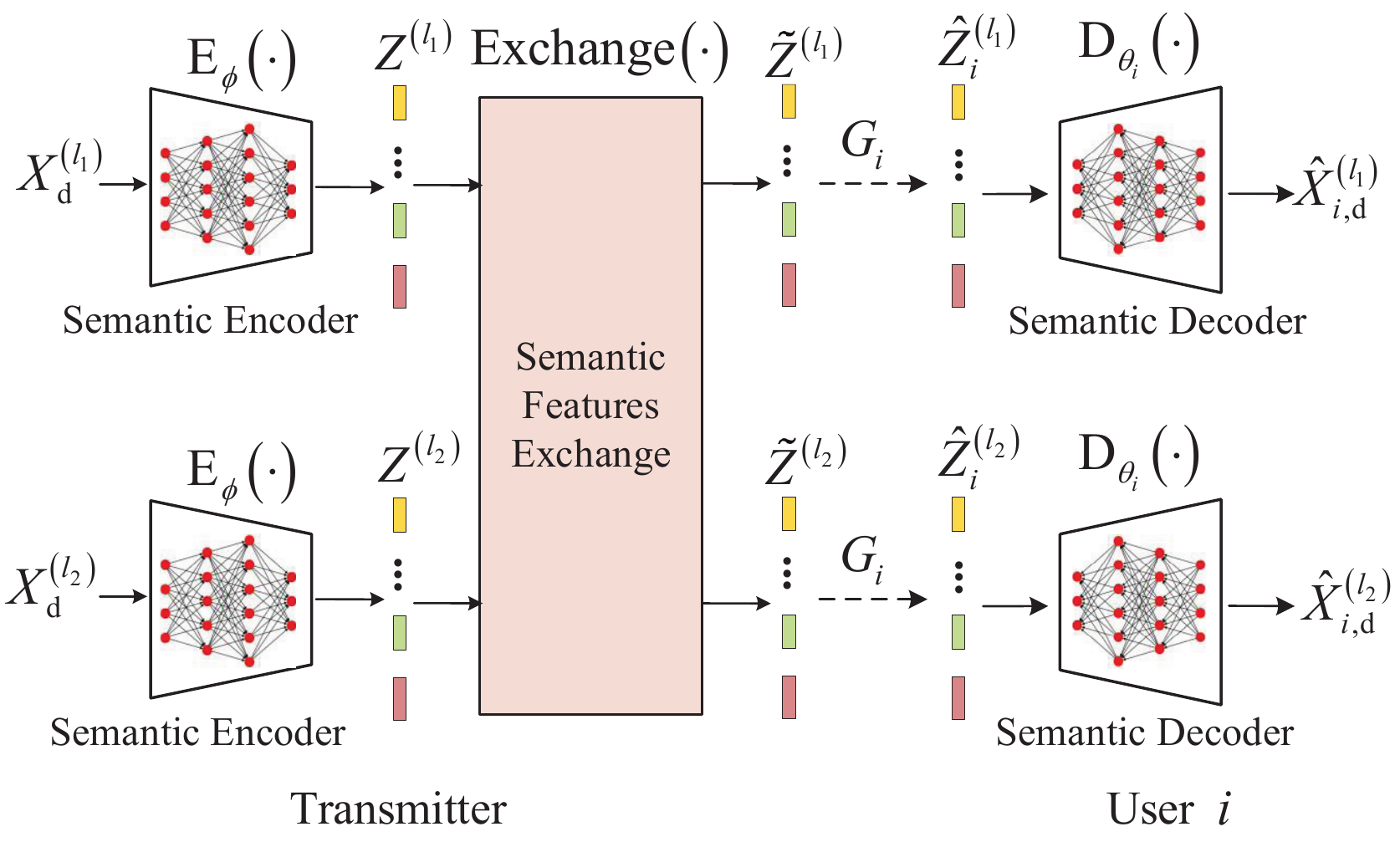}
        \vskip-0.2cm\centering {\footnotesize (b)}
    \end{minipage}
   \caption{
   (a) Self reconstruction training;
   (b) Semantic features exchange training.}
   \label{SR}
\end{figure}

To  improve the semantic features extraction ability of the network,   we further train  the   proposed semantic broadcast network by exchanging the common semantic features, and minimize  the reconstruction   loss of the common semantic features exchange.
As shown in Fig. \ref{SR} (b), we choose two input data samples ${X_{\rm{d}}^{\left( {{l_1}} \right)}}$ and ${X_{\rm{d}}^{\left( {{l_2}} \right)}}$ with   common semantic features, and
  extract  the   semantic features ${{\rm{Z}}^{\left( {{l_1}} \right)}}$ and ${{\rm{Z}}^{\left( {{l_2}} \right)}}$ through the semantic encoder ${{\rm{E}}_\phi }$   as follows
 \begin{subequations}
\begin{align}
{{\rm{Z}}^{\left( {{l_1}} \right)}} = \left[ {{\rm{z}}_1^{\left( {{l_1}} \right)},...,{\rm{z}}_K^{\left( {{l_1}} \right)}} \right]= {{\rm{E}}_\phi }\left( {X_{\rm{d}}^{\left( {{l_1}} \right)}} \right),\\
{{\rm{Z}}^{\left( {{l_2}} \right)}} =  \left[ {{\rm{z}}_1^{\left( {{l_2}} \right)},...,{\rm{z}}_K^{\left( {{l_2}} \right)}} \right] = {{\rm{E}}_\phi }\left( {X_{\rm{d}}^{\left( {{l_2}} \right)}} \right),
\end{align}
\end{subequations}
where ${l_1},{l_2} = 1,...,{L_{\rm{s}}}$. Assume that
the $j$th  semantic feature   of ${{\rm{Z}}^{\left( {{l_1}} \right)}}$ and ${{\rm{Z}}^{\left( {{l_2}} \right)}}$ are the same, i.e., ${{\rm{z}}_j^{\left( {{l_1}} \right)}}$ and ${{\rm{z}}_j^{\left( {{l_2}} \right)}}$ share the common semantic feature. Then,
we swap the   $j$th  semantic feature   of ${{\rm{Z}}^{\left( {{l_1}} \right)}}$ and ${{\rm{Z}}^{\left( {{l_2}} \right)}}$ as   follows
 \begin{subequations}
\begin{align}
\left[ \begin{array}{l}
{\widetilde {\rm{Z}}^{\left( {{l_1}} \right)}}\\
{\widetilde {\rm{Z}}^{\left( {{l_2}} \right)}}
\end{array} \right] &= {\rm{Exchange}}\left( {{{\rm{Z}}^{\left( {{l_1}} \right)}},{{\rm{Z}}^{\left( {{l_2}} \right)}},j} \right)\\
  &=  \left[ \begin{array}{l}
{\rm{z}}_1^{\left( {{l_1}} \right)},...,{\rm{z}}_j^{\left( {{l_2}} \right)},...,{\rm{z}}_K^{\left( {{l_1}} \right)}\\
{\rm{z}}_1^{\left( {{l_2}} \right)},...,{\rm{z}}_j^{\left( {{l_1}} \right)},...,{\rm{z}}_K^{\left( {{l_2}} \right)}
\end{array} \right],
\end{align}
\end{subequations}
where ${\widetilde {\rm{Z}}^{\left( {{l_1}} \right)}}$ and ${\widetilde {\rm{Z}}^{\left( {{l_2}} \right)}}$  represent the data samples after feature exchange of  ${{\rm{Z}}^{\left( {{l_1}} \right)}}$ and ${{\rm{Z}}^{\left( {{l_2}} \right)}}$ respectively,
$j = 1,...,{L_{\rm{c}}}$, and ${L_{\rm{c}}}$ denotes the total number of   the common  semantic features.
Then, the exchanged   common   semantic features ${\widetilde {\rm{Z}}^{\left( {{l_1}} \right)}}$ and ${\widetilde {\rm{Z}}^{\left( {{l_2}} \right)}}$ are   broadcasted to the $K$ users.
For User  $i$, let
$\widehat {\rm{Z}}_i^{\left( {{l_1}} \right)} $ and $\widehat {\rm{Z}}_i^{\left( {{l_2}} \right)}$,
respectively, denote the
    received semantic features of   ${\widetilde {\rm{Z}}^{\left( {{l_1}} \right)}}$ and ${\widetilde {\rm{Z}}^{\left( {{l_2}} \right)}}$, i.e.,
 \begin{subequations}
\begin{align}
&\widehat {\rm{Z}}_i^{\left( {{l_1}} \right)} = {{\rm{G}}_i}{\widetilde {\rm{Z}}^{\left( {{l_1}} \right)}} + {N_{i,{\rm{s}}}},\\
&\widehat {\rm{Z}}_i^{\left( {{l_2}} \right)} = {G_i}{\widetilde {\rm{Z}}^{\left( {{l_2}} \right)}} + {N_{i,{\rm{s}}}},
\end{align}
\end{subequations}
where $i = 1,...,K$. Furthermore, by leveraging the semantic decoder  ${{\rm{D}}_{{\theta _i}}}\left(  \cdot  \right)$, the   reconstructed data $\widehat X_{i{\rm{,d}}}^{\left( {{l_1}} \right)}$ and $\widehat X_{i{\rm{,d}}}^{\left( {{l_2}} \right)}$ are, respectively, given as
 \begin{subequations}
\begin{align}
&\widehat X_{i{\rm{,d}}}^{\left( {{l_1}} \right)} = {{\rm{D}}_{{\theta _i}}}\left( {\widehat {\rm{Z}}_i^{\left( {{l_1}} \right)}} \right),\\
&\widehat X_{i{\rm{,d}}}^{\left( {{l_2}} \right)} = {{\rm{D}}_{{\theta _i}}}\left( {\widehat {\rm{Z}}_i^{\left( {{l_2}} \right)}} \right).
\end{align}
\end{subequations}

 Finally, for the   common semantic features exchange training, the reconstruction loss of the semantic broadcast network  ${L_{{\rm{AE,2}}}}$  is given  as
 \begin{align}
{L_{{\rm{AE}},{\rm{2}}}} = \sum\limits_{i = 1}^K {\sum\limits_{\scriptstyle{l_1},{l_2} = 1,\hfill\atop
\scriptstyle{l_2} \ne {l_1}\hfill}^{{L_{\rm{c}}}} {\left( {{{\left\| {\hat X_{i,{\rm{d}}}^{\left( {{l_1}} \right)} - X_{\rm{d}}^{\left( {{l_1}} \right)}} \right\|}^2} + {{\left\| {\hat X_{i,{\rm{d}}}^{\left( {{l_2}} \right)} - X_{\rm{d}}^{\left( {{l_2}} \right)}} \right\|}^2}} \right)} } .\nonumber\end{align}

\subsubsection{Different  Semantic Features Exchange}

 In order to enhance the ability of decoupling different semantic features,
    we train
the   semantic BC  network  by exchanging the different semantic features, and minimizing  the reconstruction   loss of different semantic features exchanges.
  Specifically,
as shown in Fig. \ref{SR} (b), we choose two input data samples ${X_{\rm{d}}^{\left( {{l_1}} \right)}}$ and ${X_{\rm{d}}^{\left( {{l_3}} \right)}}$ with   different semantic features, and
  extract  the   semantic features ${{\rm{Z}}^{\left( {{l_1}} \right)}}$ and ${{\rm{Z}}^{\left( {{l_3}} \right)}}$ through the semantic encoder ${{\rm{E}}_\phi }$   as follows
 \begin{subequations}\label{Different_encoded}
\begin{align}
&{{\rm{Z}}^{\left( {{l_1}} \right)}} = {{\rm{E}}_\phi }\left( {X_{\rm{d}}^{\left( {{l_1}} \right)}} \right),\\
&{{\rm{Z}}^{\left( {{l_3}} \right)}} = {{\rm{E}}_\phi }\left( {X_{\rm{d}}^{\left( {{l_3}} \right)}}, \right)
\end{align}
\end{subequations}
where ${l_1},{l_3} = 1,...,{L_{\rm{s}}}$. Assume that
the $n$th  semantic features   of ${{\rm{Z}}^{\left( {{l_1}} \right)}}$ and ${{\rm{Z}}^{\left( {{l_3}} \right)}}$ are different, i.e., ${{\rm{z}}_j^{\left( {{l_1}} \right)}}$ and ${{\rm{z}}_n^{\left( {{l_3}} \right)}}$ are different. Then,
we swap the   $n$th  semantic feature   of ${{\rm{Z}}^{\left( {{l_1}} \right)}}$ and ${{\rm{Z}}^{\left( {{l_3}} \right)}}$ as     follows
   \begin{align}\left[ \begin{array}{l}
{\widetilde {\rm{Z}}^{\left( {{l_1}} \right)}}\\
{\widetilde {\rm{Z}}^{\left( {{l_3}} \right)}}
\end{array} \right] = {\rm{Exchange}}\left( {{{\rm{Z}}^{\left( {{l_1}} \right)}},{{\rm{Z}}^{\left( {{l_3}} \right)}},n} \right).\label{exchange3a}
\end{align}
where ${\widetilde {\rm{Z}}^{\left( {{l_1}} \right)}}$ and ${\widetilde {\rm{Z}}^{\left( {{l_3}} \right)}}$  represent the data samples after feature exchange of  ${{\rm{Z}}^{\left( {{l_1}} \right)}}$ and ${{\rm{Z}}^{\left( {{l_3}} \right)}}$ respectively,
$n = 1,...,{L_{\rm{e}}}$, and ${L_{\rm{e}}}$ denotes the total number of   the different  semantic features.

Then, the exchanged    semantic features ${\widetilde {\rm{Z}}^{\left( {{l_1}} \right)}}$ and ${\widetilde {\rm{Z}}^{\left( {{l_3}} \right)}}$  are, respectively, broadcasted to the $K$ users.
For User  $i$, let
$\widetilde {\rm{Z}}_{i,{\rm{r}}}^{\left( {{l_1}} \right)}$ and $\widetilde {\rm{Z}}_{i,{\rm{r}}}^{\left( {{l_3}} \right)}$,
respectively, denote the
    received semantic features of  ${\widetilde {\rm{Z}}^{\left( {{l_1}} \right)}}$ and ${\widetilde {\rm{Z}}^{\left( {{l_3}} \right)}}$, i.e.,
 \begin{subequations}\label{BC3}
\begin{align}
&\widetilde {\rm{Z}}_{i,{\rm{r}}}^{\left( {{l_1}} \right)} = {\alpha _i}{\widetilde {\rm{Z}}^{\left( {{l_1}} \right)}} + {N_{i,{\rm{s}}}},\\
&\widetilde {\rm{Z}}_{i,{\rm{r}}}^{\left( {{l_3}} \right)} = {\alpha _i}{\widetilde {\rm{Z}}^{\left( {{l_3}} \right)}} + {N_{i,{\rm{s}}}},
\end{align}
\end{subequations}
where   $i = 1,...,K$. Furthermore, based the semantic decoder  ${{\rm{D}}_{{\theta _i}}}\left(  \cdot  \right)$, the   reconstructed data $\widetilde X_{i,{\rm{d}}}^{\left( {{l_1}} \right)}$ and $\widetilde X_{i,{\rm{d}}}^{\left( {{l_3}} \right)}$ of User $i$ are   given as
 \begin{subequations}\label{DE3}
\begin{align}
&\widetilde X_{i,{\rm{d}}}^{\left( {{l_1}} \right)} = {{\rm{D}}_{{\theta _i}}}\left( {\widetilde {\rm{Z}}_{i,{\rm{r}}}^{\left( {{l_1}} \right)}} \right),\\
&\widetilde X_{i,{\rm{d}}}^{\left( {{l_3}} \right)} = {{\rm{D}}_{{\theta _i}}}\left( {\widetilde {\rm{Z}}_{i,{\rm{r}}}^{\left( {{l_1}} \right)}} \right).
\end{align}
\end{subequations}

Then, by taking the reconstructed  samples $\widehat X_{i{\rm{,d}}}^{\left( {{l_1}} \right)}$ and $\widehat X_{i{\rm{,d}}}^{\left( {{l_3}} \right)}$ as input data samples,
 the semantic broadcast network   repeats the semantic encoding in \eqref{Different_encoded}, different  semantic features exchange in \eqref{exchange3a}, broadcasting the  semantic features in \eqref{BC3}  and decoding semantic features in \eqref{DE3}. For brevity, we omit the details.
 Note that,
both semantic feature  exchange the  $n$th feature.  Hence,
  after two semantic feature exchanges, the different features exchanged each return to their original data samples.

After two semantic features exchanges, let ${\overline X _{i,{\rm{d}}}^{\left( {{l_1}} \right)}}$ and ${\overline X _{i,{\rm{d}}}^{\left( {{l_3}} \right)}}$ denote the final   reconstructed data, respectively,  and   the corresponding  reconstruction loss ${L_{{\rm{AE,3}}}}$ is given as
\begin{align}{L_{{\rm{AE}},{\rm{3}}}} = \sum\limits_{i = 1}^K {\sum\limits_{\scriptstyle{l_1},{l_3} = 1,\hfill\atop
\scriptstyle{l_3} \ne {l_1}\hfill}^{{L_{\rm{e}}}} {\left( {{{\left\| {\overline X_{i,{\rm{d}}}^{\left( {{l_1}} \right)} - X_{\rm{d}}^{\left( {{l_1}} \right)}} \right\|}^2} + {{\left\| {\overline X_{i,{\rm{d}}}^{\left( {{l_3}} \right)} - X_{\rm{d}}^{\left( {{l_3}} \right)}} \right\|}^2}} \right)} } .\nonumber
\end{align}
\begin{figure}[htbp]
      \centering	
      \includegraphics[width=0.8\textwidth]{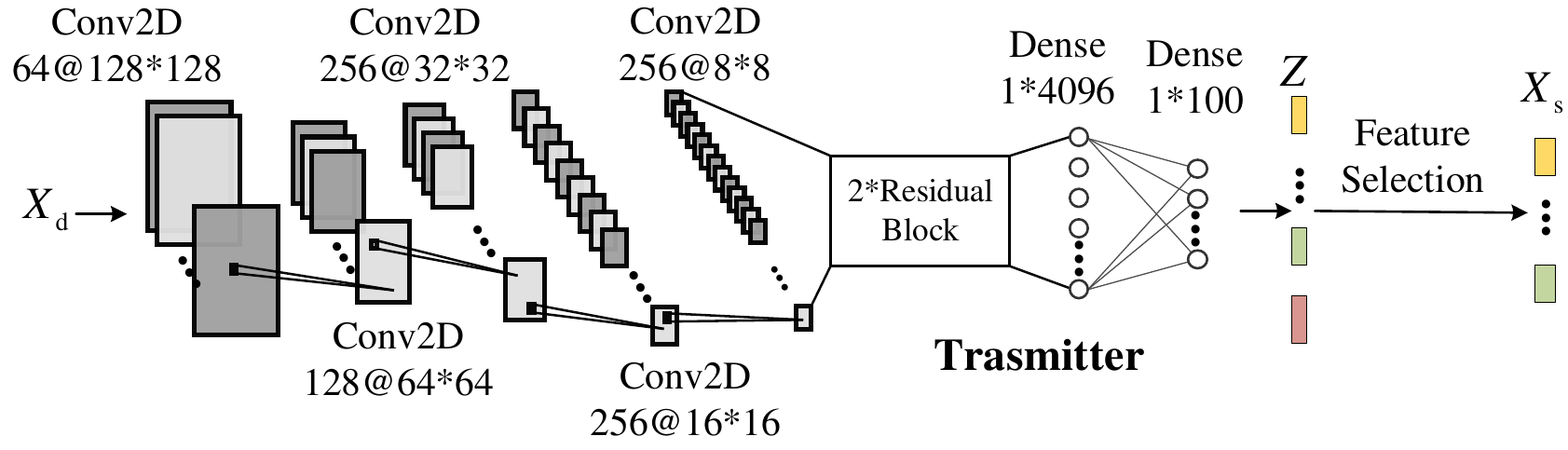}

     \vskip0.1cm \centering {\footnotesize (a)}

     \centering
     \includegraphics[width=0.8\textwidth]{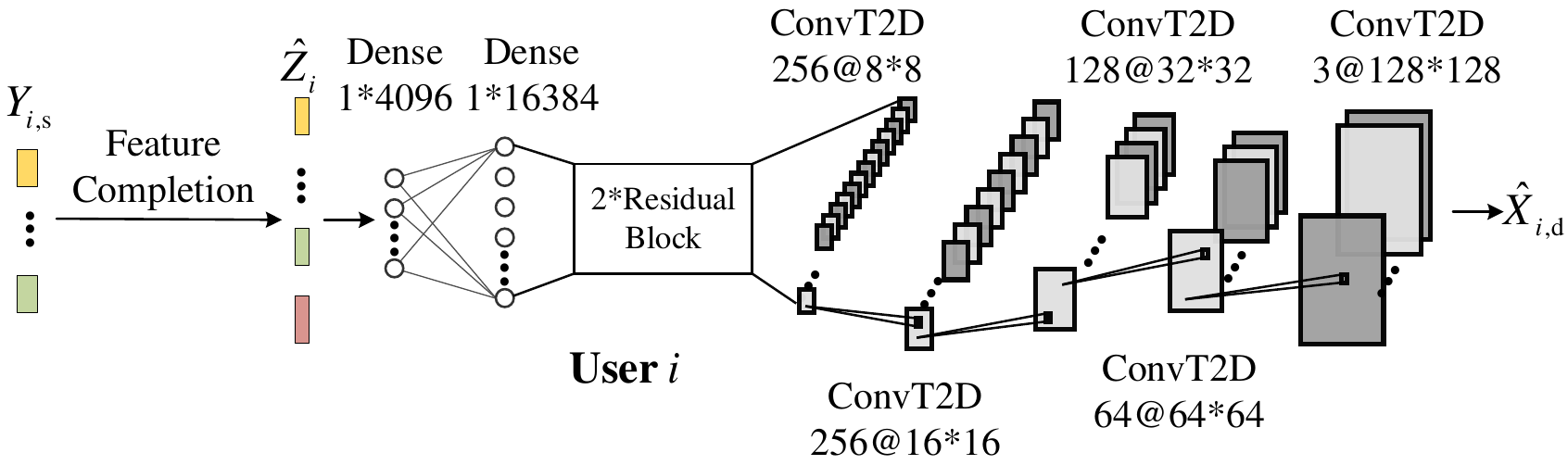}

     \vskip0.1cm\centering {\footnotesize (b)}
   \caption{
   (a) The semantic transmitter architecture ;
   (b) Semantic User $i$.}
   \label{ESM prototype1}
\end{figure}

\subsection{Proposed Semantic BC Network Architecture}

 The proposed    semantic broadcast network architecture includes a  semantic transmitter network and $K$ semantic users network, as shown in Fig. \ref{ESM prototype1}, where the notation Conv2D
  64$@$64*64 means  that the network has 64
  2-D convolutional filters with output size 64*64, Dense
  1*4096
  represents    a dense layer with 4096 neurons, and  Residual Block represents a convolution block with  Conv2D
  256$@$8*8  $\to$ Conv2D
  256$@$8*8.
  More specifically,
  the details of the proposed semantic broadcast network architecture
  are listed as follows:

  \subsubsection{The network architecture of the transmitter} ${X_{\rm{d}}}$ $\to$ Conv2D
  64$@$128*128 $\to$ Conv2D
  128$@$64*64
  $\to$ Conv2D
  256$@$32*32 $\to$ Conv2D
  256$@$16*16  $\to$ Conv2D
  256$@$8*8 $\to$ 2*Residual Block $\to$ Dense
  1*4096 $\to$
   Dense
  1*100
  $\to$ $\left\{ {{z_l}} \right\}_{l = 1}^{100}$ $\to$ ${X_{\rm{s}}}$;

  \subsubsection{The network architecture of  User $i$}
 ${Y_{i,{\rm{d}}}}$ $\to$  $\left\{ {{{\widehat z}_l}} \right\}_{l = 1}^{100}$ $\to$
  Dense
  1*4096 $\to$ Dense
  1*16384 $\to$ 2*Residual Block
  $\to$ ConvT2D
  256$@$8*8 $\to$ ConvT2D
  256$@$16*16 $\to$ ConvT2D
  128$@$32*32 $\to$ ConvT2D
  64$@$64*64 $\to$ ConvT2D
  3$@$128*128  $\to$  ${\widehat X_{i,{\rm{d}}}}$.

 	\section{Semantic BC Prototype and Implementations}

\begin{figure}[htbp]
      \centering	
      \includegraphics[width=0.7\textwidth]{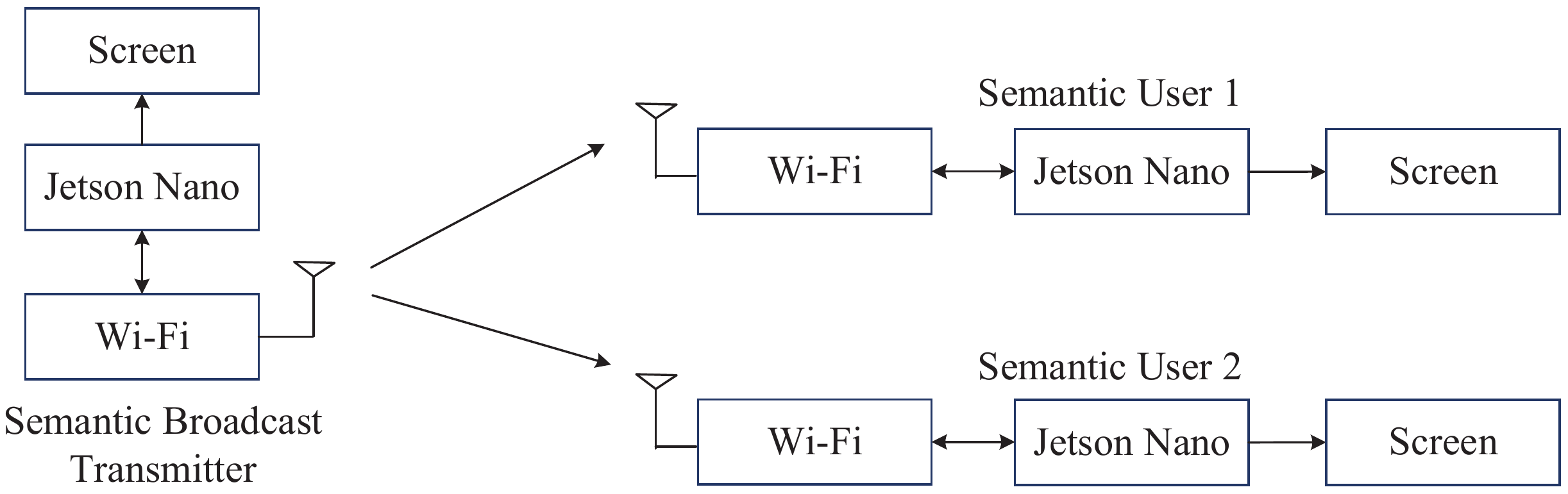}
     \vskip0.1cm\centering {\footnotesize (a)}

     \centering
     \includegraphics[width=0.7\textwidth]{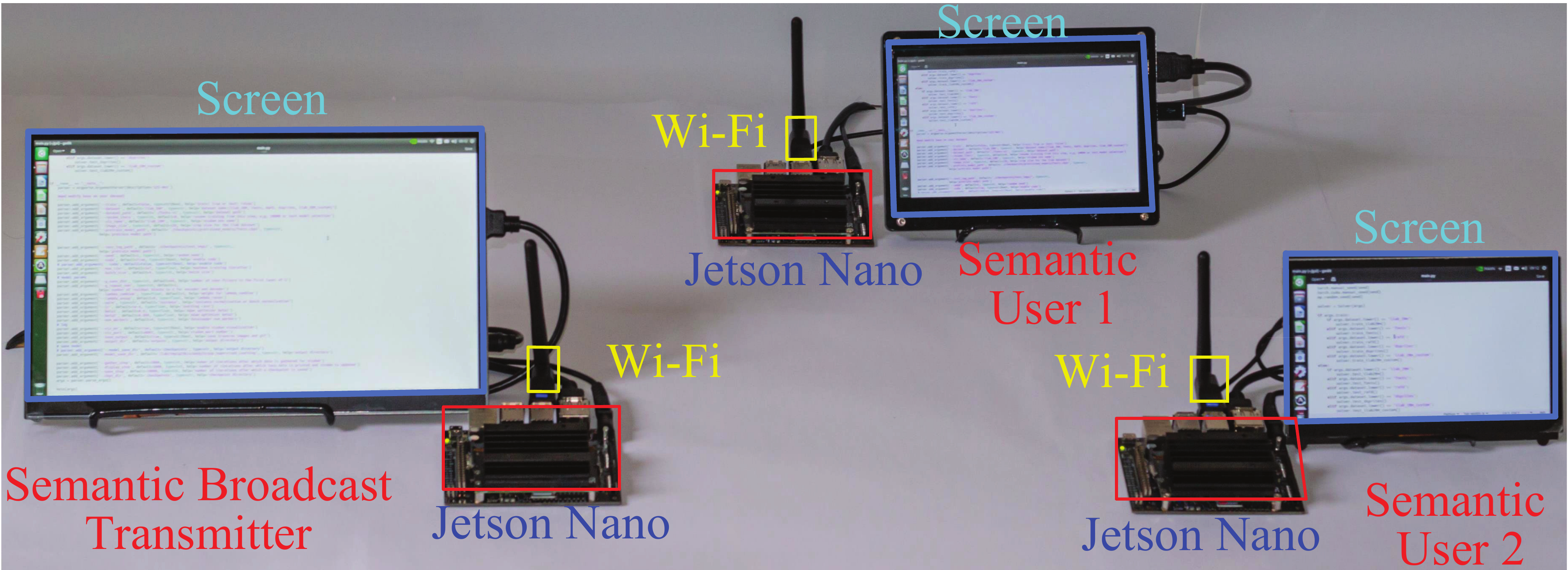}
     \vskip0.1cm\centering {\footnotesize (b)}

   \caption{
   (a) The architecture of the semantic broadcast network
  prototype with two users;
   (b) The hardware platform of the semantic broadcast network
  prototype with two users.}
   \label{ESM prototype2}
\end{figure}

The proposed semantic broadcast network  prototype architecture and the   hardware platform design
    are, respectively, shown in Fig. \ref{ESM prototype2} (a) and (b), which can be used to implement the proposed features-disentangled semantic broadcast  network
   in Fig. \ref{ESM prototype1}.
The proposed semantic broadcast network prototype includes one transmitter and two semantic
mobile users, i.e., User 1 and User 2. The trained semantic broadcast network is implemented  using three    portable Jetson Nano B01
processors, which represent the transmitter, User $1$ and User $2$. The detailed parameters of the portable Jetson Nano B01 prototype are provided in Table \ref{canshu},
which is equipped with NVIDIA Maxwell graphics processing unit (GPU) architecture with 128 NVIDIA A cores,
an ARM Cortex - A57 MPcore @quad-core CPU, Wi-Fi,  Pytorch-GPU and torch-vision software.

 Specifically, the transmitter  performs semantic  encoding on the input data,  and then does feature  selection and bit-level source and
channel encoding, and finally broadcasts the semantic feature data to User $1$ and $2$ via Wi-Fi.
With received   data through Wi-Fi,
User $1$    first performs  bit-level channel  and
 source decoding, and then does feature completion, followed by
semantic  decoding, and finally   the  decoded data is displayed.
 The operation of User $2$ is similar to that of User $1$.

%

 \begin{table}[!hbp]
	\centering
	\caption{Hardware parameters of the semantic BC network  prototype.}
	\begin{tabular}{l|l}
		\hline
			GPU  &   NVIDIA Maxwell architecture\\ & 128-NVIDIA-CUDA-core \\
		\hline
		CPU&  Quad-core Cortex-A57
 \\
  	\hline
		Memory & 4GB LPDDR4\\
				\hline
			Wi-Fi & 2.4GHz IEEE 802.11n$/$g$/$b \\
		\hline
		Screen  &  1920*1080px/800*480px    display\\
		\hline
	\end{tabular}\label{canshu}
\end{table}

 \section{ Experiments Results and Analysis}
%

In this section, the experimental performance of the proposed
feature disentangled semantic broadcast network  is    evaluated  via both the GPU simulation  and
 the hardware  prototype.
The GPU experiments in this   work have been performed on 502 GB RAM Intel Xeon Gold 6240 CPU,
 and 24 GB Nvidia GeForce 3090 GTX graphics card with Pytorch powered with CUDA 11.3.
We adopt the Adam optimizer\cite{Diederik_2014} with a batch size 16 and an initial learning rate of 0.0001.
   The experiments are performed through two standard
datasets, i.e., Fonts Dataset\cite{Ge_ICLR_2021} and ilab-20M Dataset\cite{Borji_CVPR}.

 	\subsection{ Demonstration of   Semantic BC Performance via GPU Simulation }

The robust transmission performance of the  proposed semantic BC
scheme is demonstrated   over additive  non-Gaussian distributed  noise    channels and slow Rayleigh fading channels.
 Moreover,  four BC transmission schemes are compared, i.e., the JPEG based BC  scheme, which uses the JPEG compression in digital communications,
 non-robust semantic BC scheme,   where no channel noise is added to the training process, the proposed robust semantic BC scheme with ${\rm{SNR}}_{{\rm{train}}}= 4{\rm{dB}}$, and  the proposed robust semantic BC scheme with ${\rm{SNR}}_{{\rm{train}}}= 8{\rm{dB}}$,
  where ${\rm{SNR}}_{{\rm{train}}}$ represents  the training SNR  of the   scheme.

\begin{figure}[htbp]
    \centering
    \begin{minipage}[t]{0.55\textwidth}
        \centering
        \includegraphics[width=\textwidth]{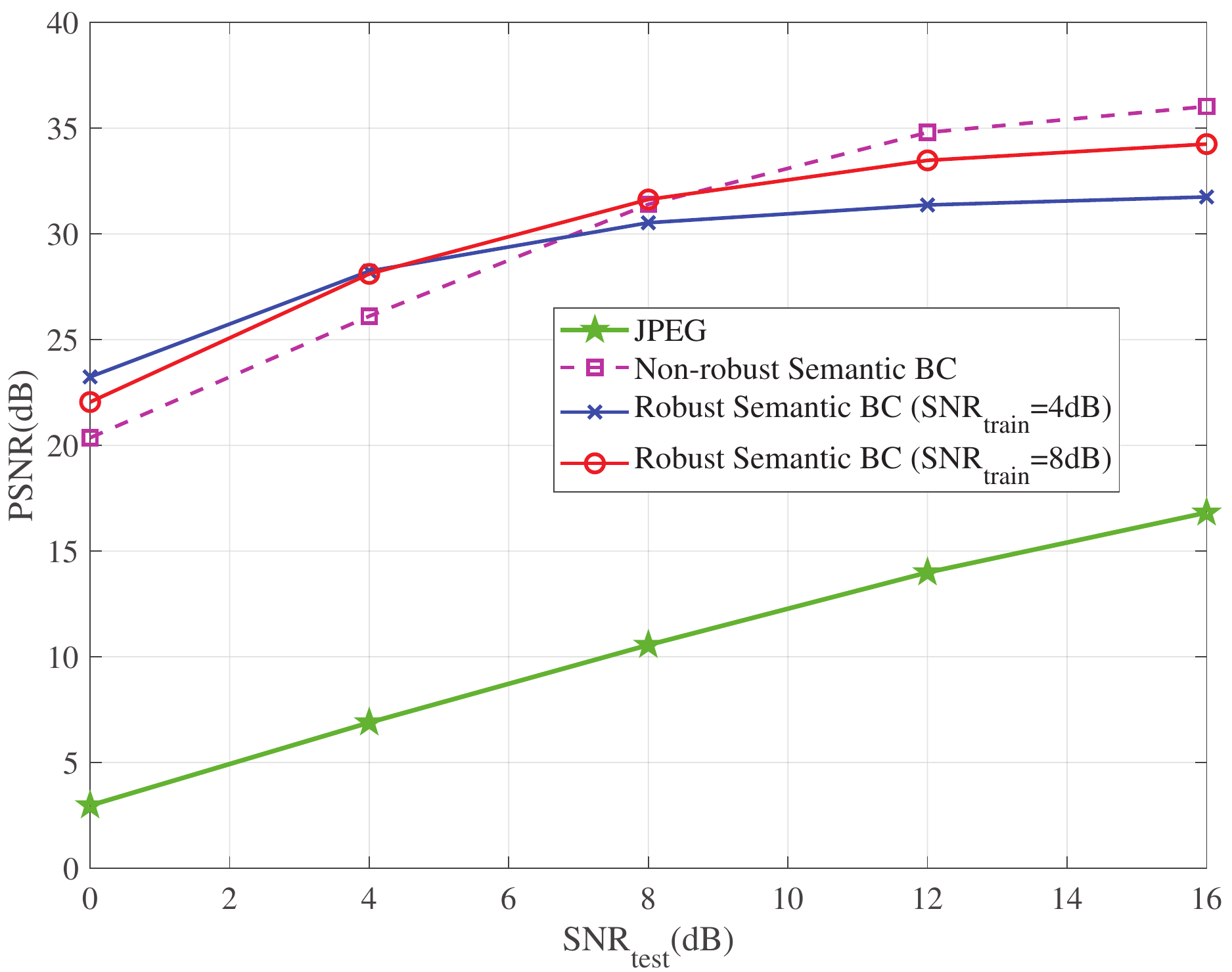}
        \vskip-0.2cm\centering {\footnotesize (a)}
    \end{minipage}
    \begin{minipage}[t]{0.55\textwidth}
        \centering
        \includegraphics[width=\textwidth]{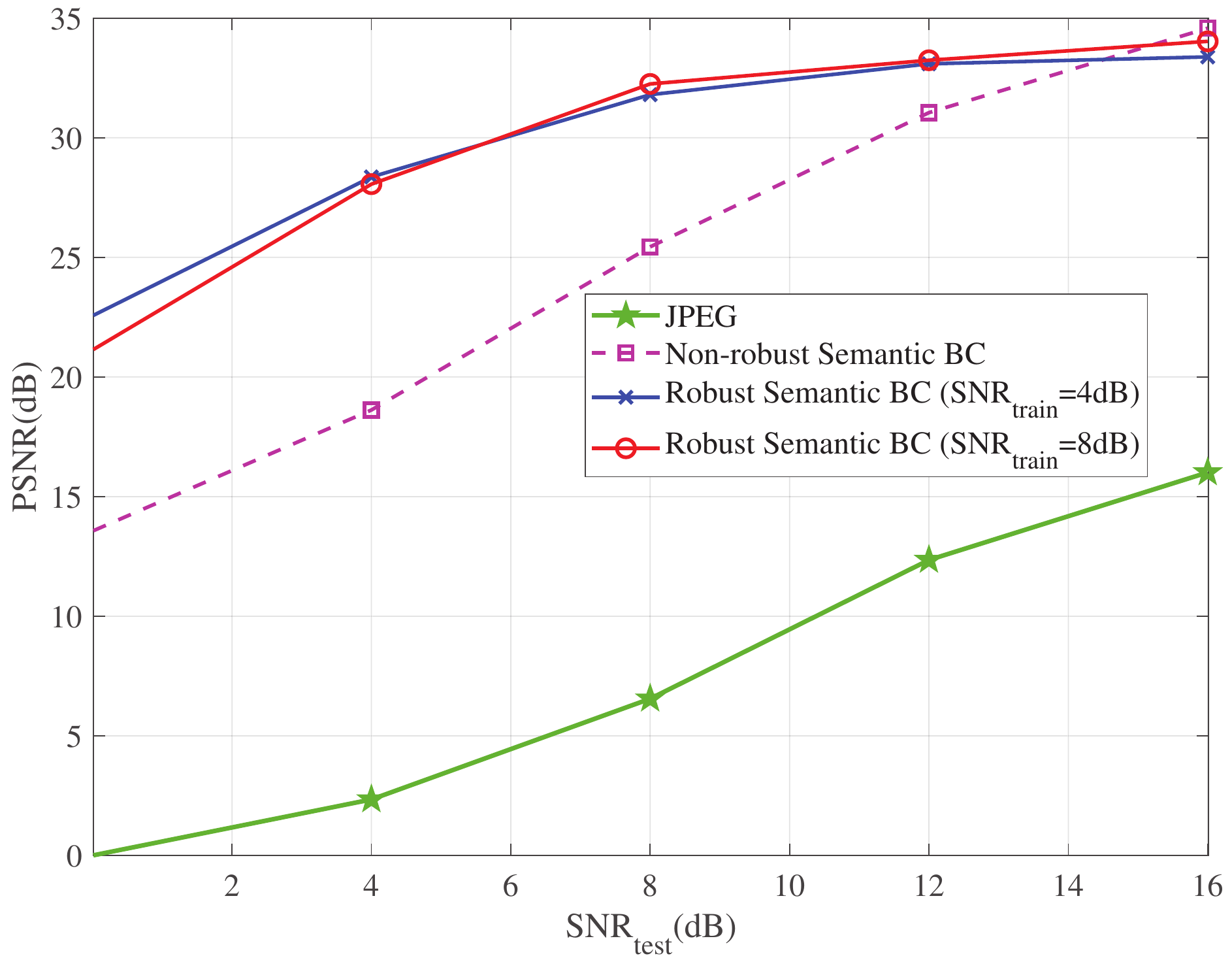}
        \vskip-0.2cm\centering {\footnotesize (b)}
    \end{minipage}
   \caption{  (a) PSNRs of    the four schemes  versus ${\rm{SNR}}_{{\rm{test}}}$ over ANGC;
   (b) PSNRs of    the four schemes  versus ${\rm{SNR}}_{{\rm{test}}}$ over   over Rayleigh fading channels.}
   \label{ANGC_Rayleigh}
\end{figure}
  Fig. \ref{ANGC_Rayleigh} (a) depicts peak signal-to-noise ratios  (PSNRs) of    the four BC schemes  versus test  SNRs ${\rm{SNR}}_{{\rm{test}}}$ over ANGC, where the PDFs  of the additive  non-Gaussian distributed  noise ${N_{1,s}}$ and $ {N_{2,s}}$  are ${p_{{N_{1,{\rm{s}}}}}}\left( x \right) = {p_{{N_{2,{\rm{s}}}}}}\left( x \right) = \frac{5}{2}\left( {{\rm{erf}}\left( {\frac{{1 - 10x}}{{10\sqrt 2 }}} \right) - {\rm{erf}}\left( {\frac{{ - 1 - 10x}}{{10\sqrt 2 }}} \right)} \right)$.
  As it can
be observed, the PSNR of  the JPEG  compression scheme is significantly lower than those of the other three schemes.  In the low SNR region (${\rm{SN}}{{\rm{R}}_{{\rm{test}}}} \le 4{\rm{dB}}$), the  PSNR of  the robust semantic BC scheme with ${\rm{SNR}}_{{\rm{train}}}= 4{\rm{dB}}$ is the highest, and  PSNR of the robust semantic BC scheme with ${\rm{SNR}}_{{\rm{train}}}= 8{\rm{dB}}$ is   higher than that of the non-robust semantic BC scheme. For the medium SNR region ($4{\rm{dB}} \le {\rm{SN}}{{\rm{R}}_{{\rm{test}}}} \le 8{\rm{dB}}$), the   PSNR of  the robust semantic BC scheme with ${\rm{SNR}}_{{\rm{train}}}= 8{\rm{dB}}$ is the highest, which verifies the robustness of the proposed schemes especially for low and medium SNR regions.
 For the high SNR region (${\rm{SN}}{{\rm{R}}_{{\rm{test}}}} \ge 8{\rm{dB}}$), the   PSNR of  the non-robust semantic BC scheme is the highest. This is because the   influence of noise can be neglected in the high SNR region.

 Fig. \ref{ANGC_Rayleigh} (b) shows   PSNRs  of    the four BC schemes    versus test   SNRs (${\rm{SNR}}_{{\rm{test}}}$) over slow Rayleigh fading channels, where    ${G_1} \sim \mathcal{N}\left( {0,1} \right)$ and ${G_2} \sim \mathcal{N}\left( {0,2} \right)$.
 Similar to PSNR  performance in Fig. \ref{ANGC_Rayleigh} (a),  JPEG compression has the worst performance  among the four schemes.
  When ${\rm{SN}}{{\rm{R}}_{{\rm{test}}}} \le 15{\rm{dB}}$, the PSNRs of the proposed robust semantic BC schemes with ${\rm{SNR}}_{{\rm{train}}}= 4{\rm{dB}}$ and ${\rm{SNR}}_{{\rm{train}}}= 8{\rm{dB}}$ are       higher than that of the non-robust semantic BC scheme, which   also demonstrates the robustness of the proposed schemes. While for ${\rm{SN}}{{\rm{R}}_{{\rm{test}}}} \ge 15{\rm{dB}}$,
  the   PSNR of  the non-robust semantic BC scheme is the highest, because the noise effect  can be neglected.

\begin{table*}[htbp]
	\small
		 \caption{Performance comparison of the  semantic BC network over ANGC}
			\centering
			\setlength{\tabcolsep}{1mm}
			\begin{tabular}{ | c | c | c | c | c | c | c|c|}
				\hline
			\tabincell{c}{	Input data\\ ${X_{\rm{d}}}$} & Users & \tabincell{c}{Intended \\ feature} & \tabincell{c}{Knowledge \\ base} & JPEG &\tabincell{c}{Non-robust \\semantic BC } &\tabincell{c}{Robust\\ semantic BC \\${\rm{SNR}}_{{\rm{train}}}= 4{\rm{dB}}$}&\tabincell{c}{Robust \\ semantic BC \\${\rm{SNR}}_{{\rm{train}}}= 8{\rm{dB}}$ } \\ \hline
				\multirow{2} * {
					\begin{minipage}[b]{0.09\columnwidth}
						\centering
						\raisebox{-1\height}{\includegraphics[width=\linewidth]{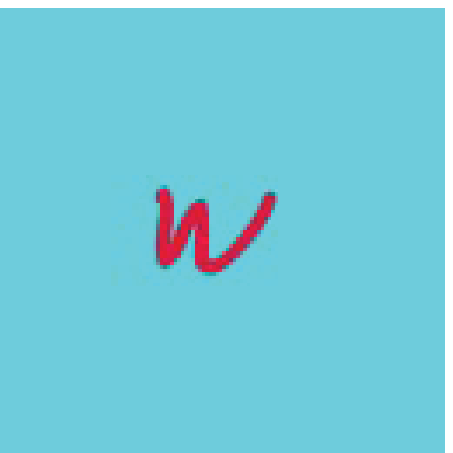}}
				\end{minipage}}&
				\tabincell{c}{User 1 \\ ${\rm{SNR}}_{{\rm{test}}}= 4{\rm{dB}}$ }& Content &\begin{minipage}[b]{0.09\columnwidth}
				\centering
				\raisebox{-.5\height}{\includegraphics[width=\linewidth]{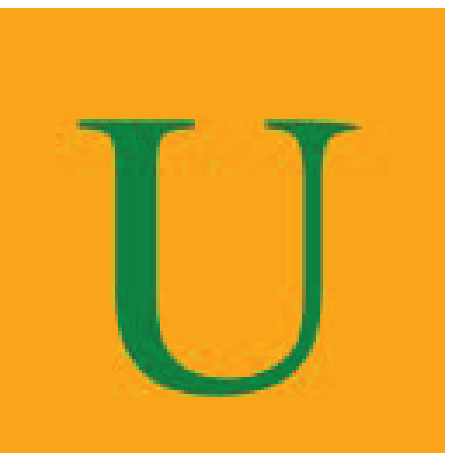}}
			    \end{minipage}&\begin{minipage}[b]{0.09\columnwidth}
					\centering
					 \raisebox{-.5\height}{\includegraphics[width=\linewidth]{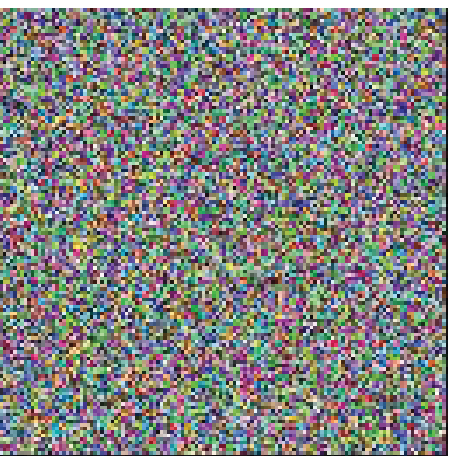}}
				\end{minipage}
				&\begin{minipage}[b]{0.09\columnwidth}
					\centering
					 \raisebox{-.5\height}{\includegraphics[width=\linewidth]{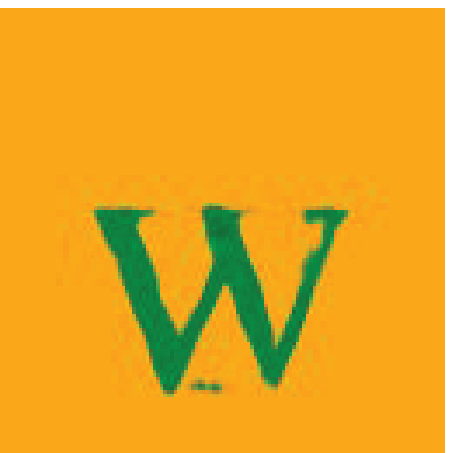}}
				\end{minipage}
				&\begin{minipage}[b]{0.09\columnwidth}
					
					\centering
					 \raisebox{-.5\height}{\includegraphics[width=\linewidth]{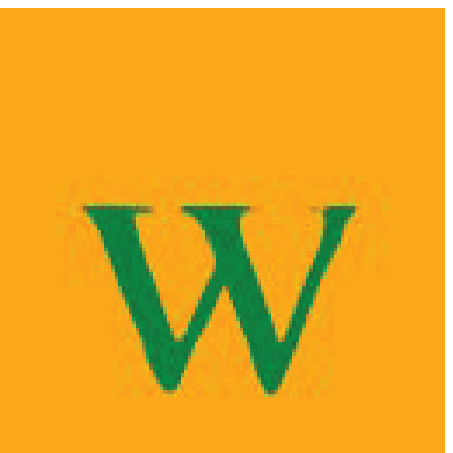}}
				\end{minipage}
				&\begin{minipage}[b]{0.09\columnwidth}
					\centering
					 \raisebox{-.5\height}{\includegraphics[width=\linewidth]{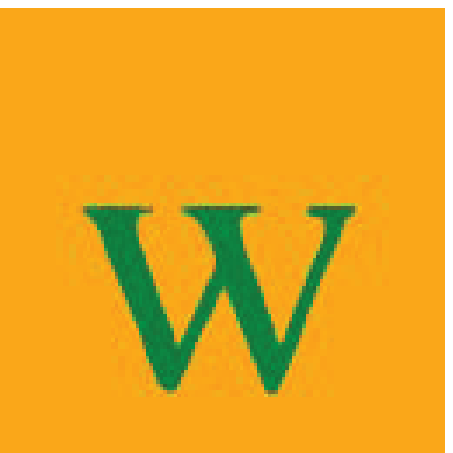}}
				\end{minipage}

				\\  \cline{2-8}&
				
				\tabincell{c}{User 2 \\ ${\rm{SNR}}_{{\rm{test}}}= 8{\rm{dB}}$ }&  \tabincell{c}{Font \\ color}&\begin{minipage}[b]{0.09\columnwidth}
					\centering
					\raisebox{-.5\height}{\includegraphics[width=\linewidth]{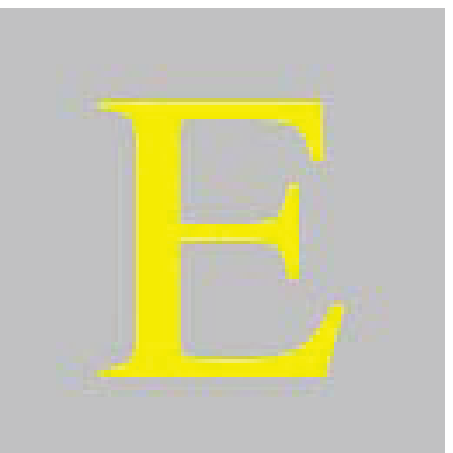}}
				\end{minipage} &\begin{minipage}[b]{0.09\columnwidth}
					\centering
					 \raisebox{-.5\height}{\includegraphics[width=\linewidth]{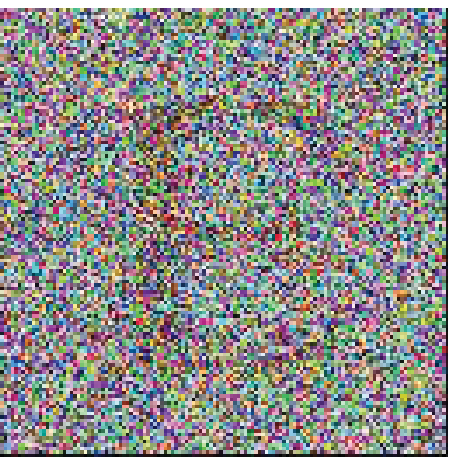}}
				\end{minipage}
				&\begin{minipage}[b]{0.09\columnwidth}
					\centering
					 \raisebox{-.5\height}{\includegraphics[width=\linewidth]{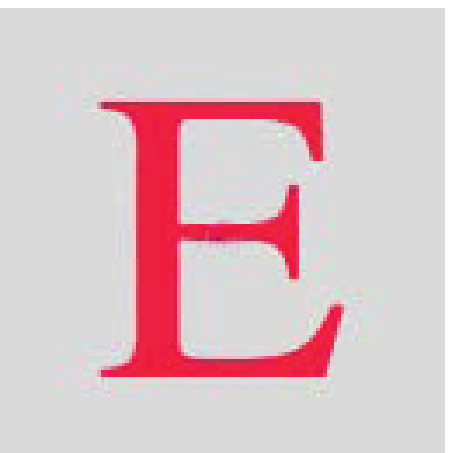}}
				\end{minipage}
				&\begin{minipage}[b]{0.09\columnwidth}
					\centering
					 \raisebox{-.5\height}{\includegraphics[width=\linewidth]{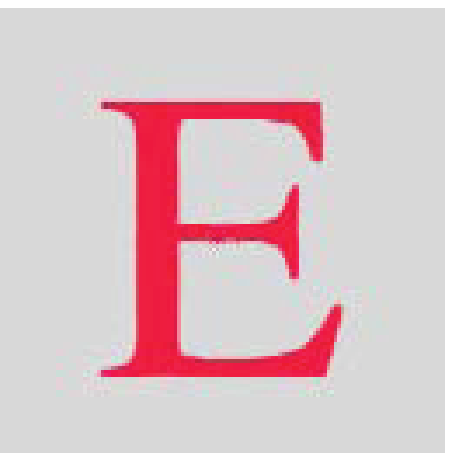}}
				\end{minipage}
				&\begin{minipage}[b]{0.09\columnwidth}
					\centering
					 \raisebox{-.5\height}{\includegraphics[width=\linewidth]{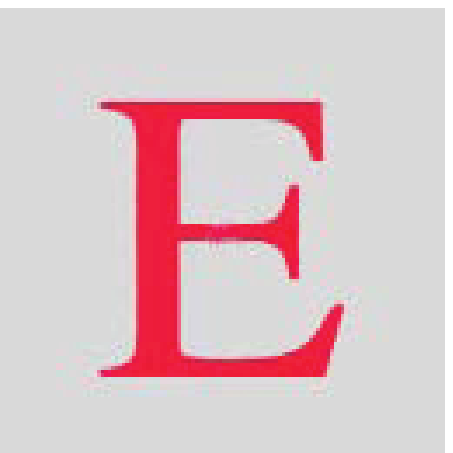}}
				\end{minipage}
				
				\\ \hline
				\multirow{2} * {
					\begin{minipage}[b]{0.09\columnwidth}
						\centering
						\raisebox{-1\height}{\includegraphics[width=\linewidth]{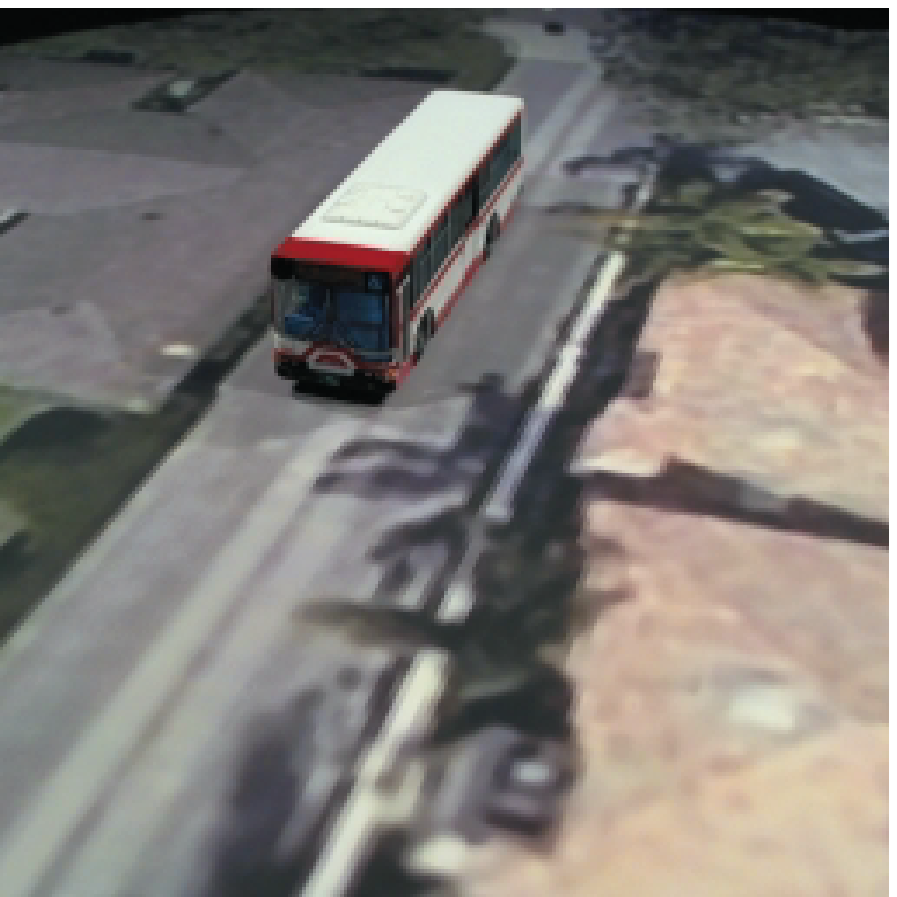}}
				\end{minipage}}&
				\tabincell{c}{User 1 \\ ${\rm{SNR}}_{{\rm{test}}}= 4{\rm{dB}}$ }& Content  &\begin{minipage}[b]{0.09\columnwidth}
				\centering
				\raisebox{-.5\height}{\includegraphics[width=\linewidth]{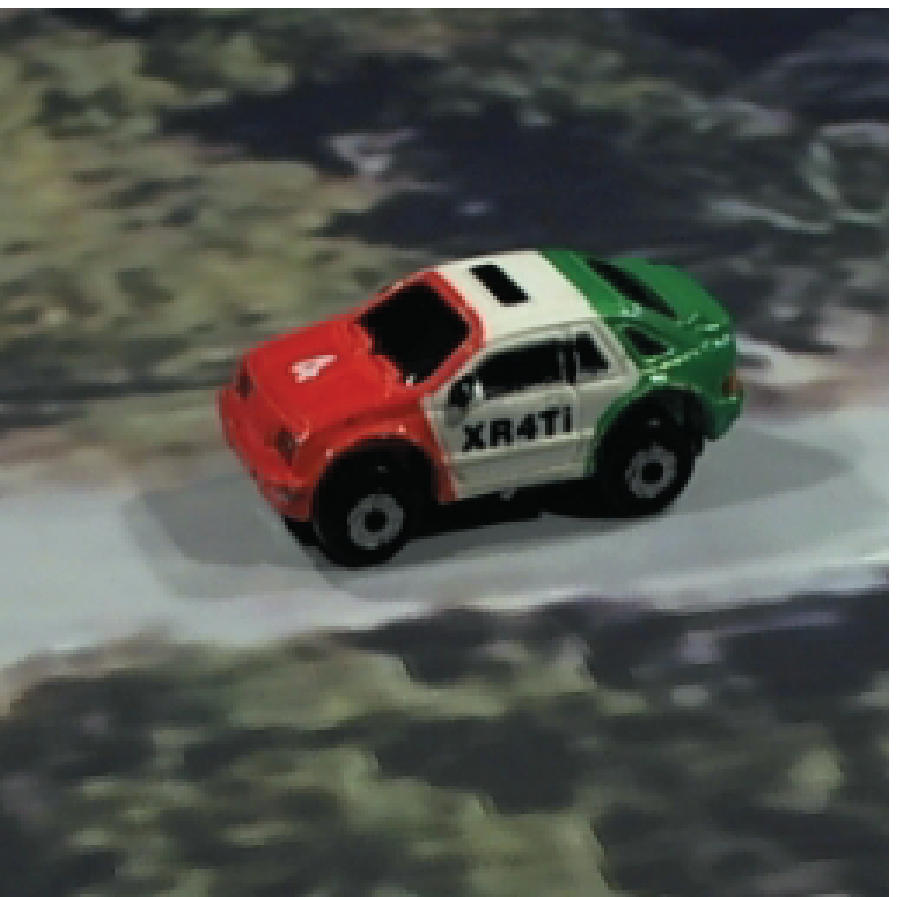}}
			\end{minipage}&\begin{minipage}[b]{0.09\columnwidth}
					\centering
					 \raisebox{-.5\height}{\includegraphics[width=\linewidth]{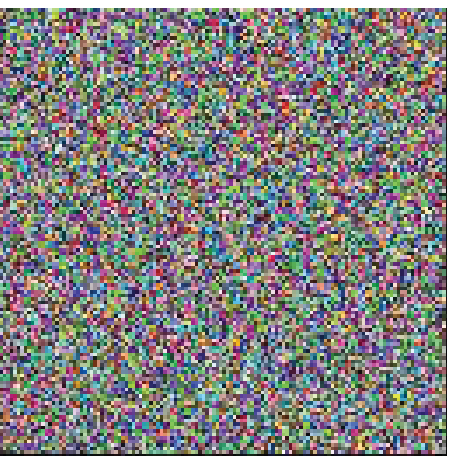}}
				\end{minipage}
				&\begin{minipage}[b]{0.09\columnwidth}
					\centering
					 \raisebox{-.5\height}{\includegraphics[width=\linewidth]{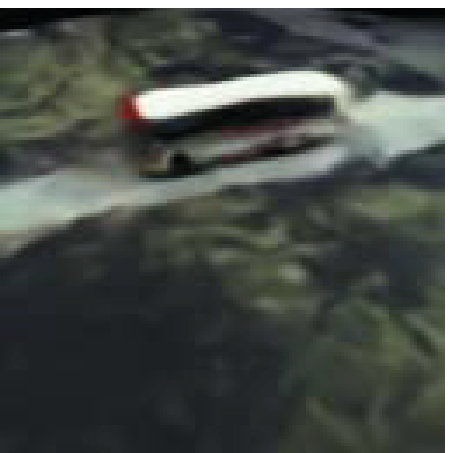}}
				\end{minipage}
				&\begin{minipage}[b]{0.09\columnwidth}
					\centering
					 \raisebox{-.5\height}{\includegraphics[width=\linewidth]{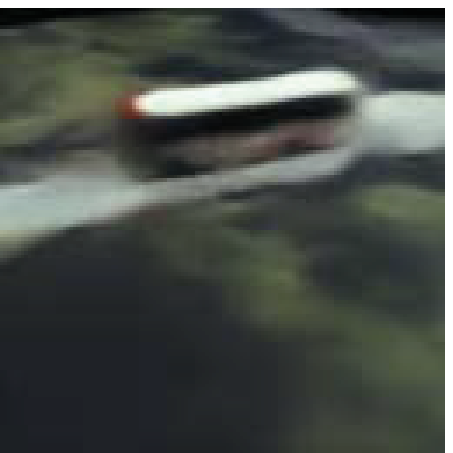}}
				\end{minipage}
				&\begin{minipage}[b]{0.09\columnwidth}
					\centering
					 \raisebox{-.5\height}{\includegraphics[width=\linewidth]{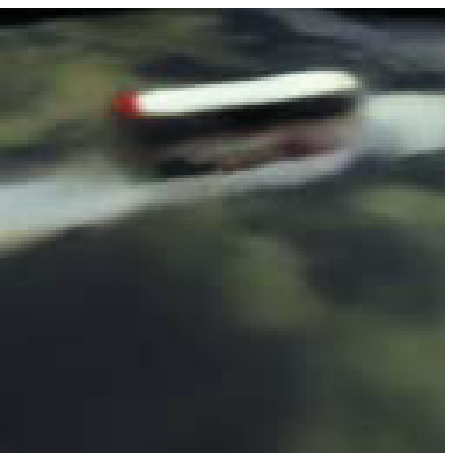}}
				\end{minipage}
				
				\\ \cline{2-8}&
				
				\tabincell{c}{User 2 \\ ${\rm{SNR}}_{{\rm{test}}}= 8{\rm{dB}}$ }& Pose &\begin{minipage}[b]{0.09\columnwidth}
				\centering
				\raisebox{-.5\height}{\includegraphics[width=\linewidth]{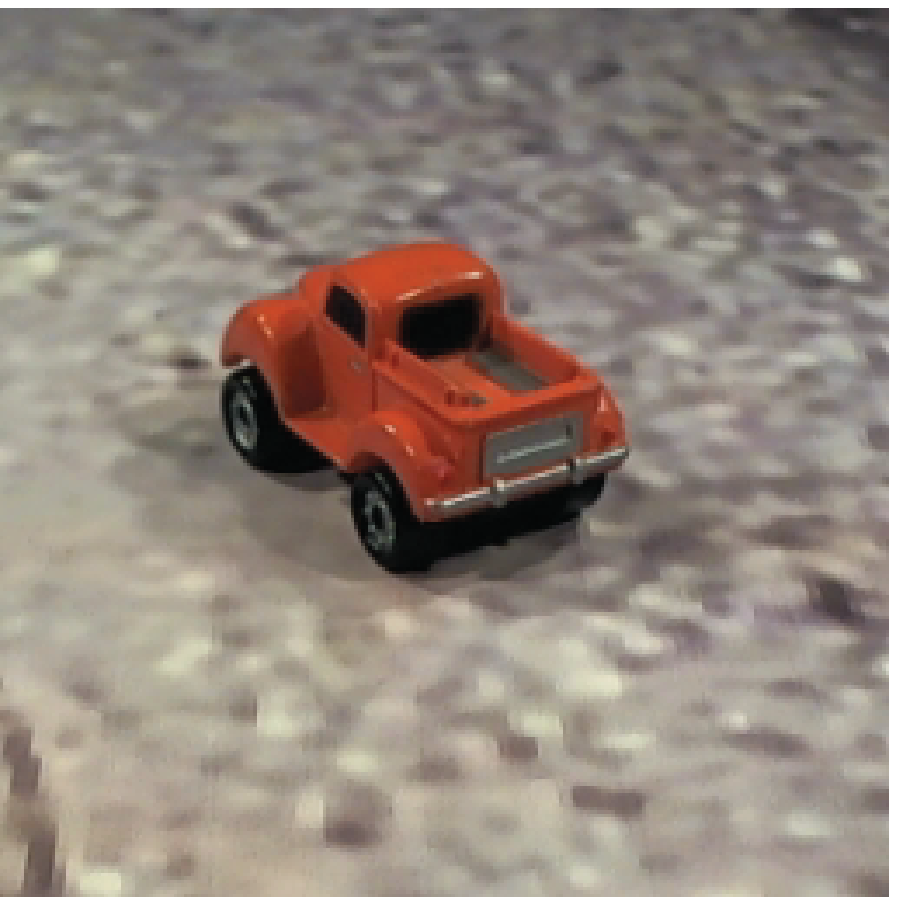}}
			\end{minipage}&\begin{minipage}[b]{0.09\columnwidth}
					\centering
					 \raisebox{-.5\height}{\includegraphics[width=\linewidth]{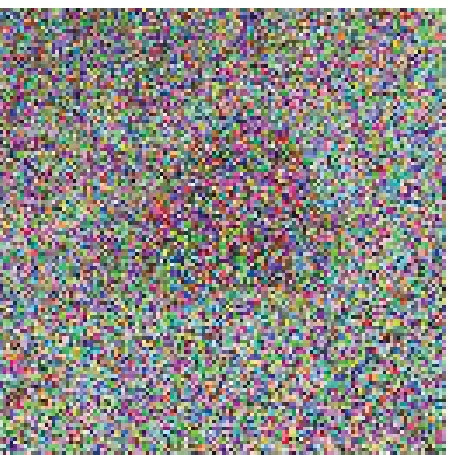}}
				\end{minipage}
				&\begin{minipage}[b]{0.09\columnwidth}
					\centering
					 \raisebox{-.5\height}{\includegraphics[width=\linewidth]{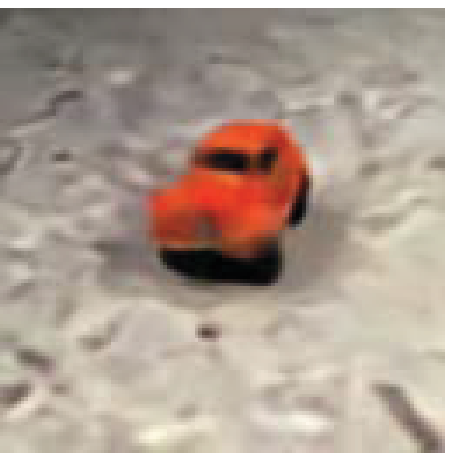}}
				\end{minipage}
				&\begin{minipage}[b]{0.09\columnwidth}
					\centering
					 \raisebox{-.5\height}{\includegraphics[width=\linewidth]{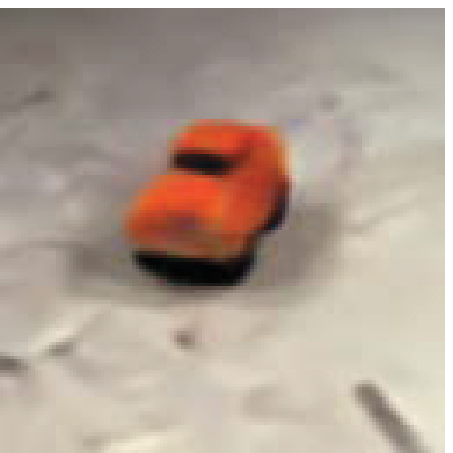}}
				\end{minipage}
				&\begin{minipage}[b]{0.09\columnwidth}
					\centering
					 \raisebox{-.5\height}{\includegraphics[width=\linewidth]{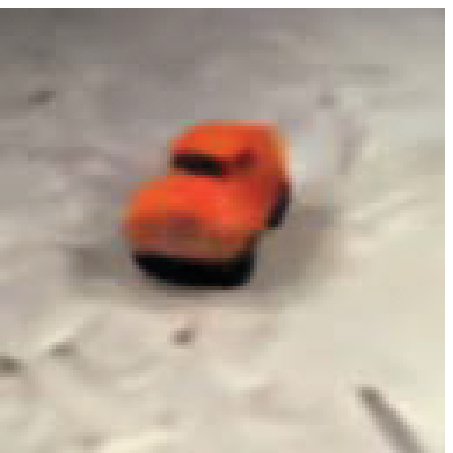}}
				\end{minipage}
				
				\\ \hline
				
			\end{tabular}\label{Transmission_ANGC}
		\end{table*}

\begin{table*}[htbp]
	\small
			\caption{Performance comparison of the  semantic BC network over Rayleigh fading channels}
			\centering
			\setlength{\tabcolsep}{1mm}
			\begin{tabular}{ | c | c | c | c | c | c | c|c|}
				\hline
		\tabincell{c}{	Input data\\ ${X_{\rm{d}}}$} & Users & \tabincell{c}{Interest \\ feature} &  \tabincell{c}{Knowledge \\ base} & JPEG &\tabincell{c}{Non-robust \\semantic BC } &\tabincell{c}{Robust \\semantic BC\\${\rm{SNR}}_{{\rm{train}}}= 4{\rm{dB}}$}&\tabincell{c}{Robust\\ semantic BC \\${\rm{SNR}}_{{\rm{train}}}= 8{\rm{dB}}$ } \\ \hline
				\multirow{2} * {
				\begin{minipage}[b]{0.09\columnwidth}
					\centering
					\raisebox{-1\height}{\includegraphics[width=\linewidth]{Data-z.eps}}
				\end{minipage}}&
				\tabincell{c}{User 1 \\ ${\rm{SNR}}_{{\rm{test}}}= 4{\rm{dB}}$ }& Content &\begin{minipage}[b]{0.09\columnwidth}
					\centering
					\raisebox{-.5\height}{\includegraphics[width=\linewidth]{KB1-z.eps}}
				\end{minipage}&\begin{minipage}[b]{0.09\columnwidth}
					\centering
					\raisebox{-.5\height}{\includegraphics[width=\linewidth]{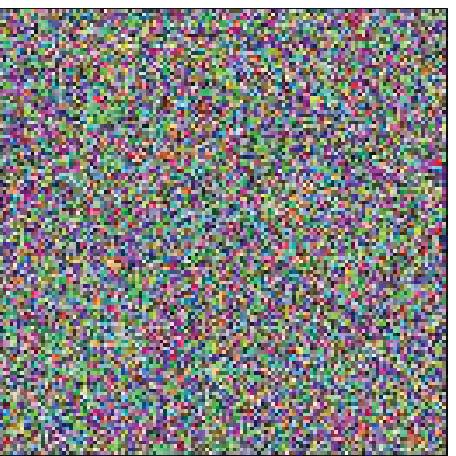}}
				\end{minipage}
				&\begin{minipage}[b]{0.09\columnwidth}
					\centering
					\raisebox{-.5\height}{\includegraphics[width=\linewidth]{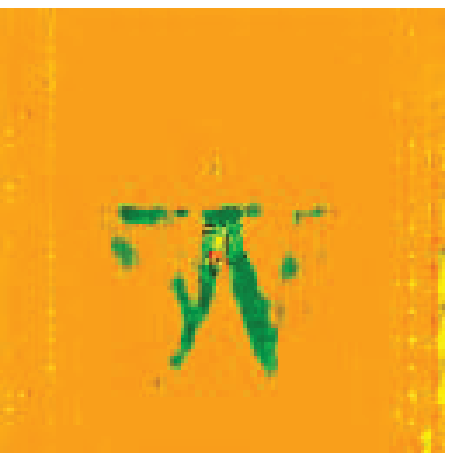}}
				\end{minipage}
				&\begin{minipage}[b]{0.09\columnwidth}
					
					\centering
					\raisebox{-.5\height}{\includegraphics[width=\linewidth]{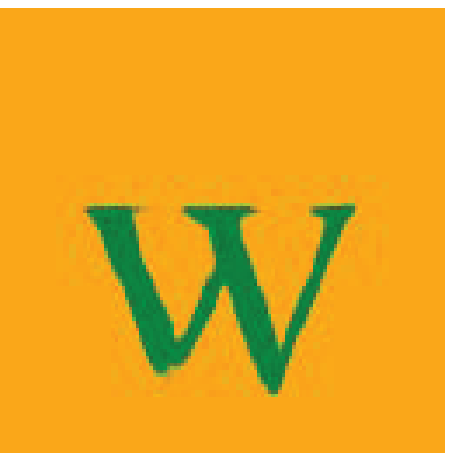}}
				\end{minipage}
				&\begin{minipage}[b]{0.09\columnwidth}
					\centering
					\raisebox{-.5\height}{\includegraphics[width=\linewidth]{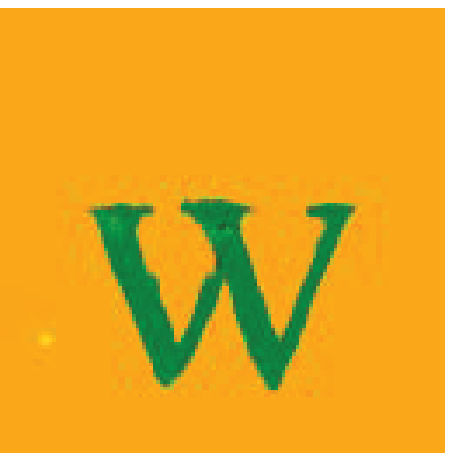}}
				\end{minipage}

				\\ \cline{2-8}
				&
				\tabincell{c}{User 2 \\ ${\rm{SNR}}_{{\rm{test}}}= 8{\rm{dB}}$ }&  \tabincell{c}{Font \\ color} &\begin{minipage}[b]{0.09\columnwidth}
					\centering
					\raisebox{-.5\height}{\includegraphics[width=\linewidth]{KB2-z.eps}}
				\end{minipage}&\begin{minipage}[b]{0.09\columnwidth}
					\centering
					\raisebox{-.5\height}{\includegraphics[width=\linewidth]{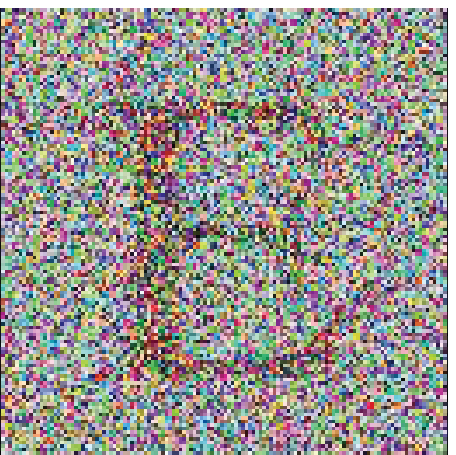}}
				\end{minipage}
				&\begin{minipage}[b]{0.09\columnwidth}
					\centering
					\raisebox{-.5\height}{\includegraphics[width=\linewidth]{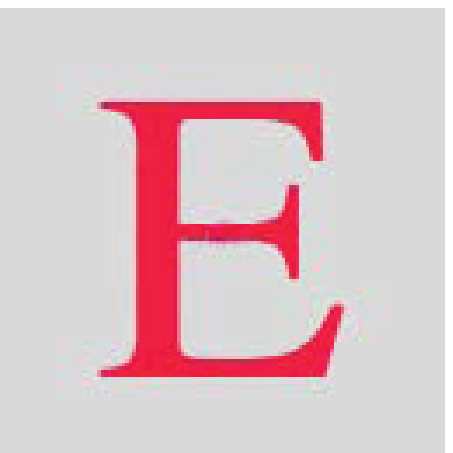}}
				\end{minipage}
				&\begin{minipage}[b]{0.09\columnwidth}
					\centering
					\raisebox{-.5\height}{\includegraphics[width=\linewidth]{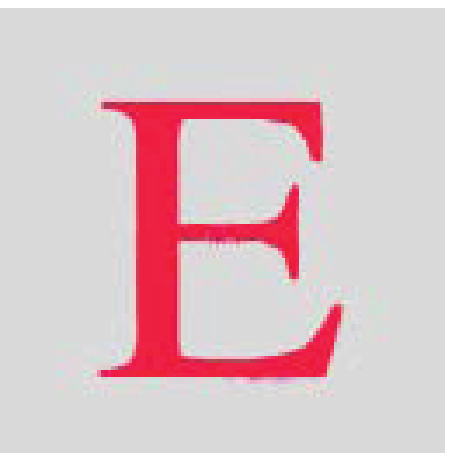}}
				\end{minipage}
				&\begin{minipage}[b]{0.09\columnwidth}
					\centering
					\raisebox{-.5\height}{\includegraphics[width=\linewidth]{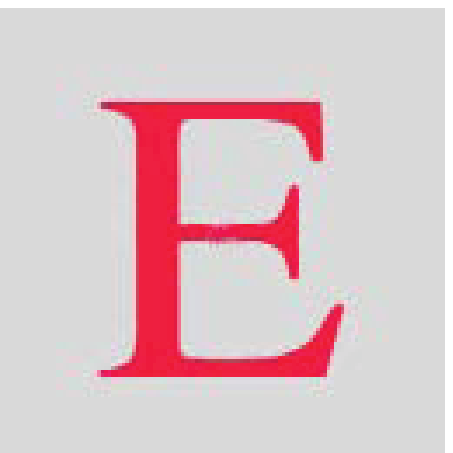}}
				\end{minipage}
				
				\\ \hline
				\multirow{2} * {
				\begin{minipage}[b]{0.09\columnwidth}
					\centering
					\raisebox{-1\height}{\includegraphics[width=\linewidth]{Data-c.eps}}
				\end{minipage}}&
				\tabincell{c}{User 1 \\ ${\rm{SNR}}_{{\rm{test}}}= 4{\rm{dB}}$ }& Content  &\begin{minipage}[b]{0.09\columnwidth}
					\centering
					\raisebox{-.5\height}{\includegraphics[width=\linewidth]{KB1-c.eps}}
				\end{minipage}&\begin{minipage}[b]{0.09\columnwidth}
					\centering
					\raisebox{-.5\height}{\includegraphics[width=\linewidth]{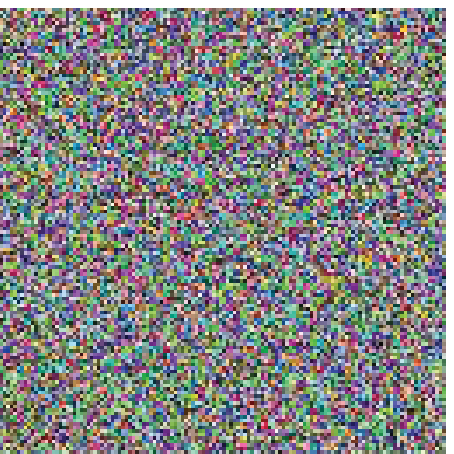}}
				\end{minipage}
				&\begin{minipage}[b]{0.09\columnwidth}
					\centering
					\raisebox{-.5\height}{\includegraphics[width=\linewidth]{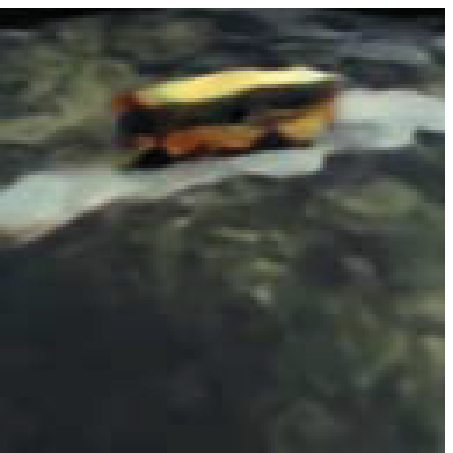}}
				\end{minipage}
				&\begin{minipage}[b]{0.09\columnwidth}
					\centering
					\raisebox{-.5\height}{\includegraphics[width=\linewidth]{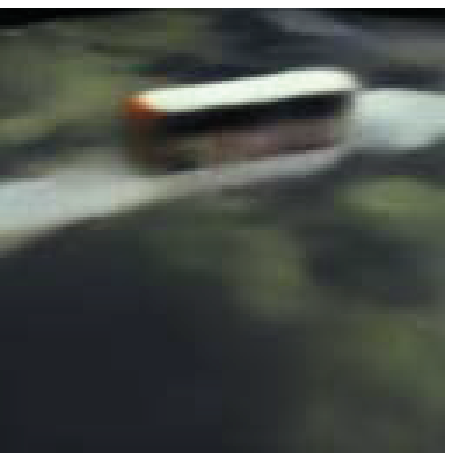}}
				\end{minipage}
				&\begin{minipage}[b]{0.09\columnwidth}
					\centering
					\raisebox{-.5\height}{\includegraphics[width=\linewidth]{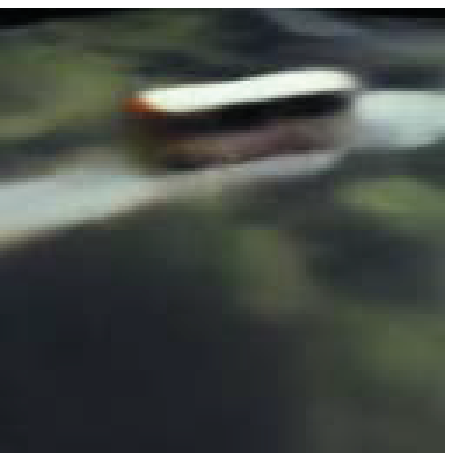}}
				\end{minipage}
				
				\\ \cline{2-8}
				&
				\tabincell{c}{User 2 \\ ${\rm{SNR}}_{{\rm{test}}}= 8{\rm{dB}}$ }& Pose &\begin{minipage}[b]{0.09\columnwidth}
					\centering
					\raisebox{-.5\height}{\includegraphics[width=\linewidth]{KB2-c.eps}}
				\end{minipage}&\begin{minipage}[b]{0.09\columnwidth}
					\centering
					\raisebox{-.5\height}{\includegraphics[width=\linewidth]{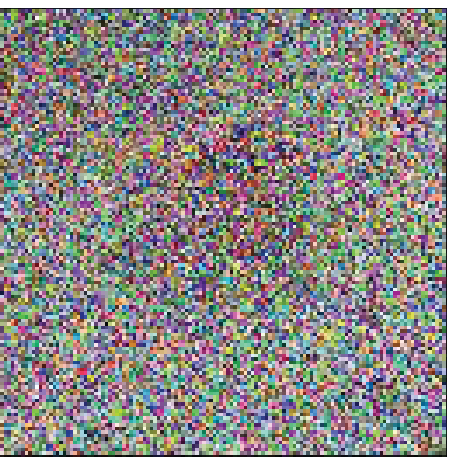}}
				\end{minipage}
				&\begin{minipage}[b]{0.09\columnwidth}
					\centering
					\raisebox{-.5\height}{\includegraphics[width=\linewidth]{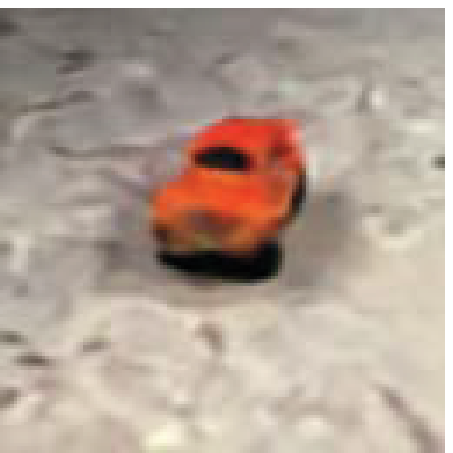}}
				\end{minipage}
				&\begin{minipage}[b]{0.09\columnwidth}
					\centering
					\raisebox{-.5\height}{\includegraphics[width=\linewidth]{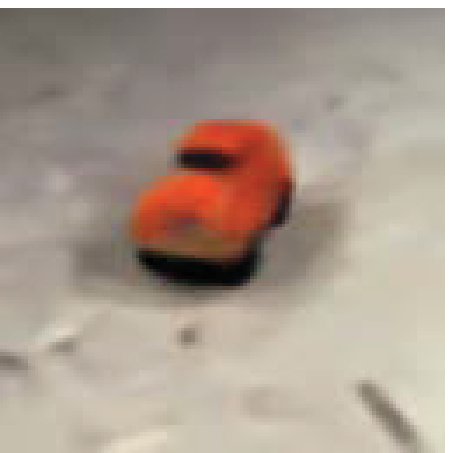}}
				\end{minipage}
				&\begin{minipage}[b]{0.09\columnwidth}
					\centering
					\raisebox{-.5\height}{\includegraphics[width=\linewidth]{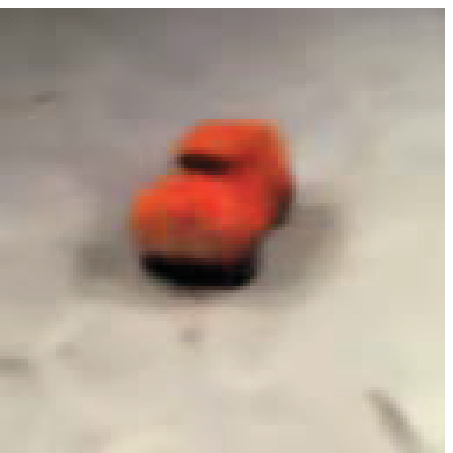}}
				\end{minipage}
				
				\\ \hline
				
			\end{tabular}\label{Transmission_Rayleigh}
		\end{table*}

		Table \ref{Transmission_ANGC} shows  performance comparison of the  semantic BC network    with two users, i.e., User 1 with ${\rm{SNR}}_{{\rm{test}}}= 4{\rm{dB}}$ and User 2 with ${\rm{SNR}}_{{\rm{test}}}= 8{\rm{dB}}$, over the ANGC, where the intended features and the knowledge bases of the two users are different. Specifically,
as shown in the  first two rows,
 the input data   is a red   lowercase letter $w$ on blue background image, and    the third   column shows that
  the intended features of  Users $1$ and $2$ are content and font color, respectively.
   The fourth   column shows that
   the knowledge bases of  Users $1$ and $2$ are a green   uppercase letter $U$ on yellow background, and a yellow
 uppercase letter $E$ on gray background, respectively.
The fifth column   shows   the poor transmission performance  of the
JPEG compression scheme, where the decoded images cannot be recognized.
 The sixth to eighth columns   are the reconstructed images of the non-robust semantic BC scheme, the robust semantic BC scheme with ${\rm{SNR}}_{{\rm{train}}}= 4{\rm{dB}}$, and  the   robust semantic BC scheme with ${\rm{SNR}}_{{\rm{train}}}= 8{\rm{dB}}$, respectively. The reconstructed data  of User 1 is a green lowercase letter $w$ on   yellow background, and the reconstructed data  of User 2 is a red uppercase letter $E$      on   gray background, which   verifies the effectiveness of the feature selection and feature completion  of the proposed semantic BC network.

Moreover,
 the input data  of  the  last two rows in Table \ref{Transmission_ANGC}
    is a
    white bus with lower-left-pose  image, and    the third   column shows that
  the intended features of  Users $1$ and $2$ are content and pose, respectively.
   The fourth   column shows that
   the knowledge bases of  Users $1$ and $2$ are a left-pose car  with red-white-green color,
      and a  red pickup truck with upper-right-pose, respectively. Similarly,   the fifth column   shows    the decoded images of  the JPEG compression scheme cannot be recognized.
 The sixth to eighth columns  show that the reconstructed data  of Users 1 and $2$   includes  intended features and the features of the users' knowledge bases,  where only the intended features are transmitted. This  also  verifies the effectiveness of the feature selection and feature completion  of the proposed semantic BC network.		
		
		Table \ref{Transmission_Rayleigh} shows  performance comparison of the four BC schemes over the Rayleigh fading channels. Similar to 	Table \ref{Transmission_ANGC}, Table \ref{Transmission_Rayleigh} demonstrates the effectiveness of the feature selection and feature completion  of the proposed semantic BC network. Moreover, the   sixth  column   shows that the reconstructed data  of  User 1 with the non-robust semantic BC scheme is   blurry, while the reconstructed data  of   the  robust semantic BC schemes are clear, which     validates the added value of the robust semantic BC design.

 	\subsection{  Demonstration of  Semantic BC Performance Via Prototype Experiment }

\begin{table*}[htbp]
	\caption{ Feature selection performance of the   semantic BC network prototype}
	\centering
	\begin{tabular}{ | c | c | c | c | c |}
		\hline
	\tabincell{c}{	Input data\\ ${X_{\rm{d}}}$} & Users & \tabincell{c}{Intended \\ feature} &  \tabincell{c}{Knowledge \\ base} & \tabincell{c}{Reconstructed\\ data  ${\widehat X_{i,{\rm{d}}}}$ \\ (Jetson Nano)}  \\ \hline
		\multirow{2} * {
			\begin{minipage}[b]{0.1\columnwidth}
				\centering
				\raisebox{-1\height}{\includegraphics[width=\linewidth]{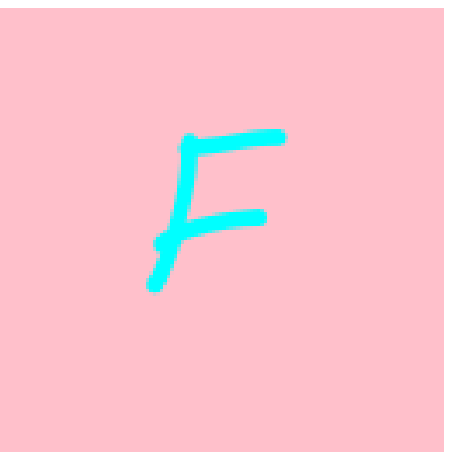}}
		\end{minipage}}&
		\tabincell{c}{User 1}& Content &\begin{minipage}[b]{0.1\columnwidth}
			\centering
			\raisebox{-.5\height}{\includegraphics[width=\linewidth]{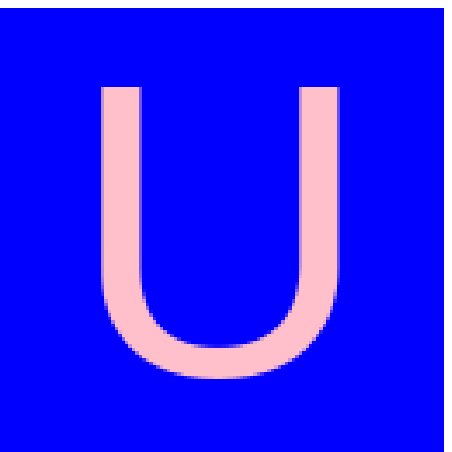}}
		\end{minipage}&\begin{minipage}[b]{0.1\columnwidth}
			\centering
			\raisebox{-.5\height}{\includegraphics[width=\linewidth]{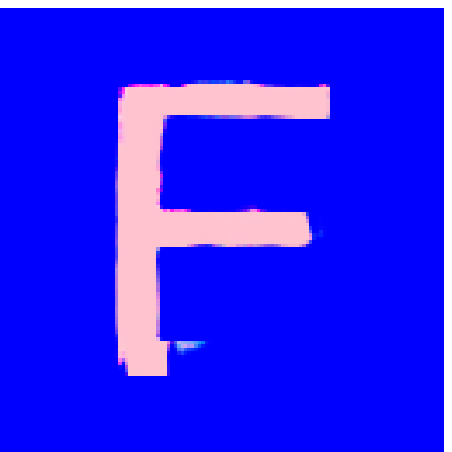}}
		\end{minipage}

		\\ \cline{2-5}
		&
		\tabincell{c}{User 2  }&  \tabincell{c}{Style} &\begin{minipage}[b]{0.1\columnwidth}
			\centering
			\raisebox{-.5\height}{\includegraphics[width=\linewidth]{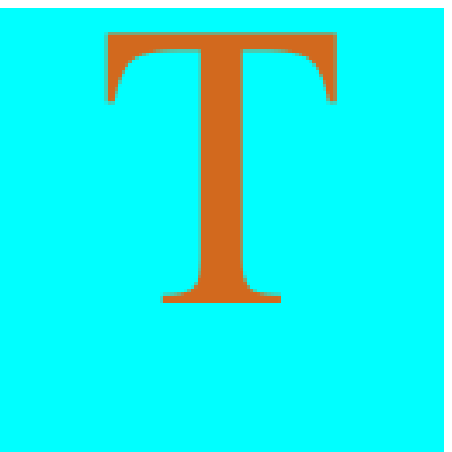}}
		\end{minipage}&\begin{minipage}[b]{0.1\columnwidth}
			\centering
			\raisebox{-.5\height}{\includegraphics[width=\linewidth]{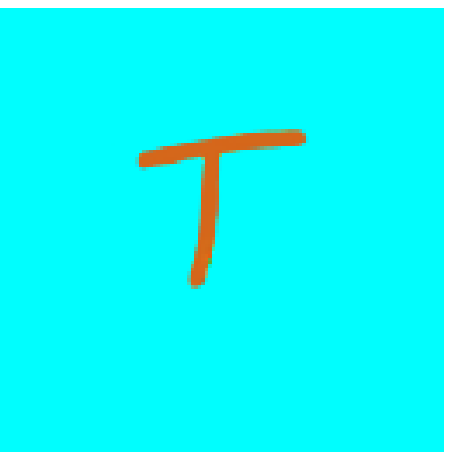}}
		\end{minipage}

		\\ \hline
			
		\multirow{2} * {
			\begin{minipage}[b]{0.1\columnwidth}
				\centering
				\raisebox{-1\height}{\includegraphics[width=\linewidth]{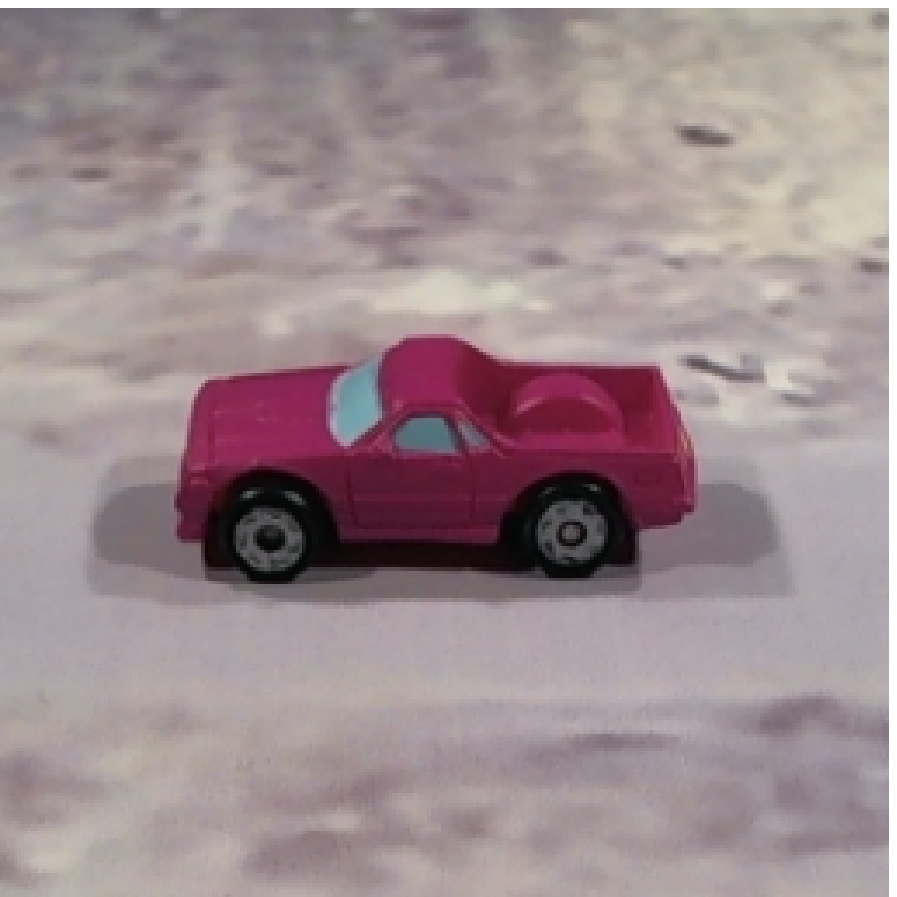}}
		\end{minipage}}&
		\tabincell{c}{User 1}& Content  &\begin{minipage}[b]{0.1\columnwidth}
			\centering
			\raisebox{-.5\height}{\includegraphics[width=\linewidth]{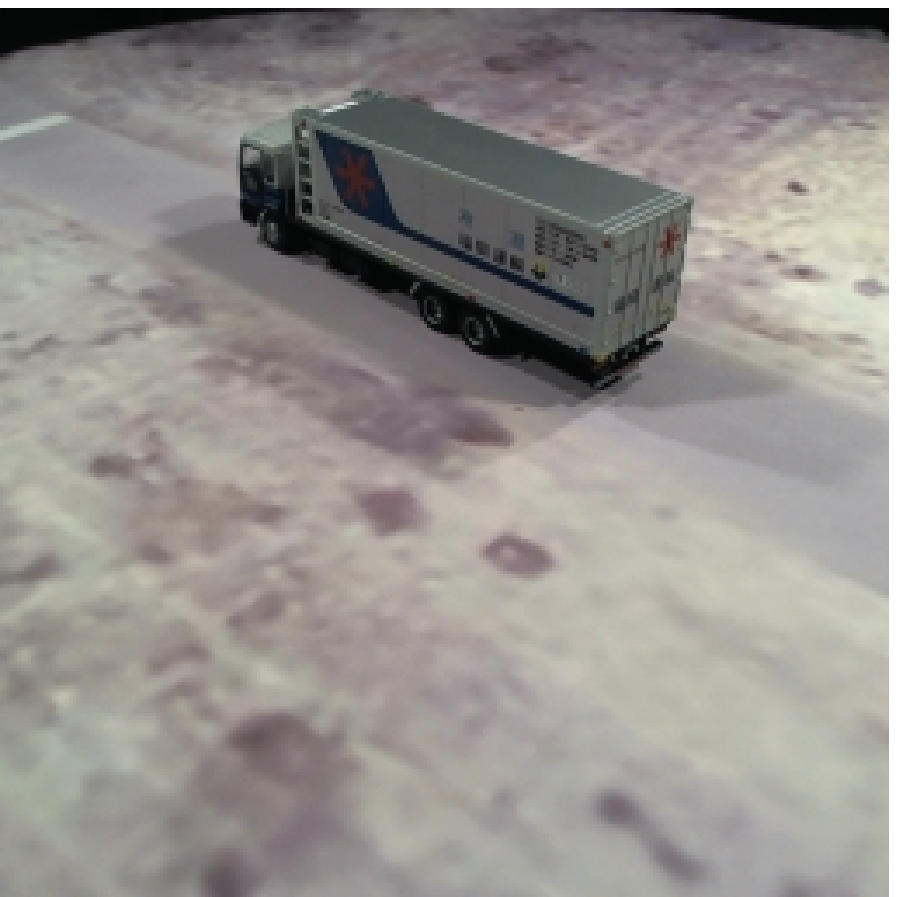}}
		\end{minipage}&\begin{minipage}[b]{0.1\columnwidth}
			\centering
			\raisebox{-.5\height}{\includegraphics[width=\linewidth]{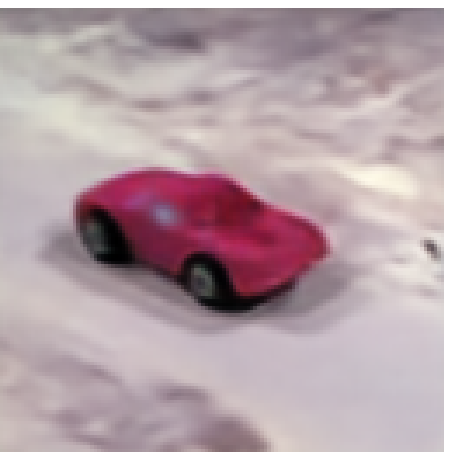}}
		\end{minipage}

		\\ \cline{2-5}
		&
		\tabincell{c}{User 2 }& Pose &\begin{minipage}[b]{0.1\columnwidth}
			\centering
			\raisebox{-.5\height}{\includegraphics[width=\linewidth]{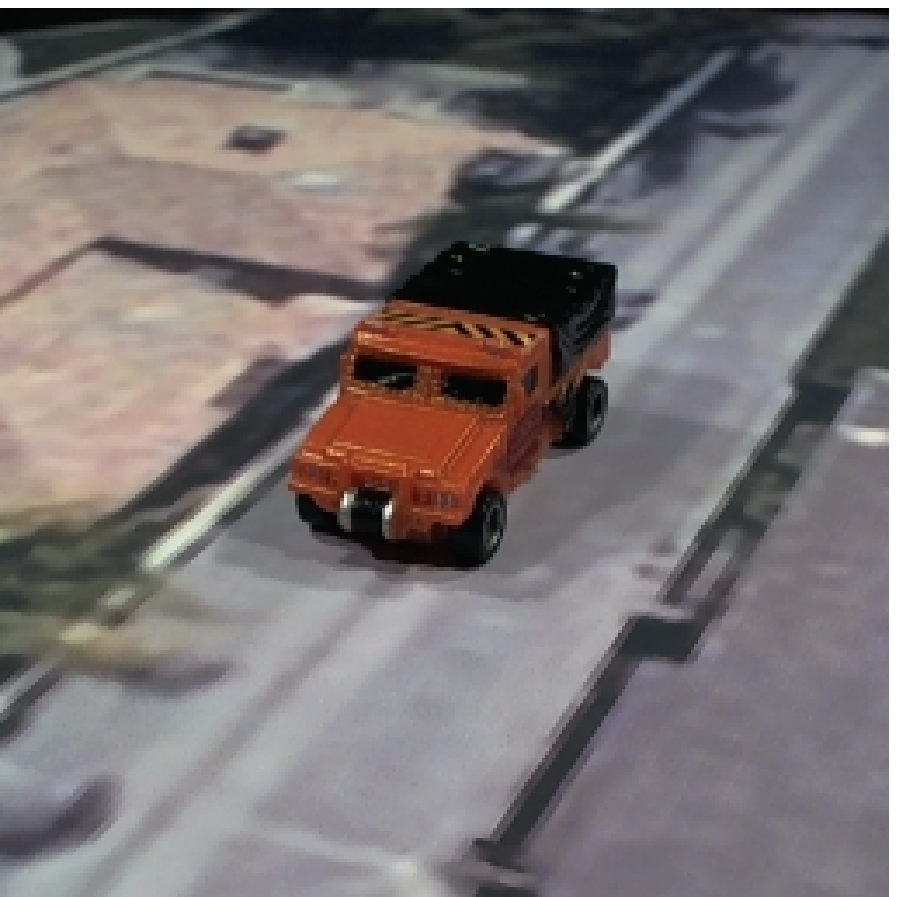}}
		\end{minipage}&\begin{minipage}[b]{0.1\columnwidth}
			\centering
			\raisebox{-.5\height}{\includegraphics[width=\linewidth]{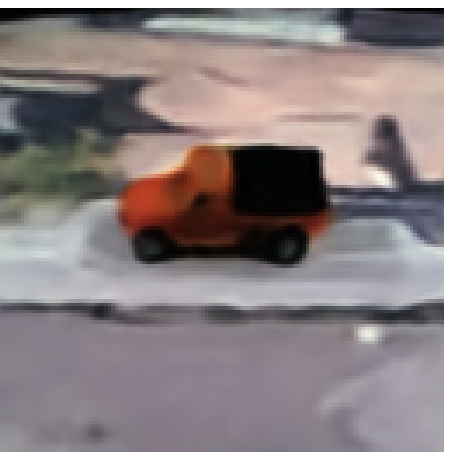}}
		\end{minipage}

		\\ \hline
		
	\end{tabular}\label{Transmission_prototype}
\end{table*}

The performance of the features-disentangled   semantic BC networks
  are
   demonstrated   via the proposed semantic BC prototype, which was designed in  Section VI,

 Table \ref{Transmission_prototype} shows feature selection performance  of the  semantic BC network,
 where
  the intended features and the knowledge bases of   Users $1$ and $2$   are different. Specifically,
  the input data  of  the  first two rows is  a small size blue letter $F$ in italics,
     and    the third   column shows that
  the intended features of  Users $1$ and $2$ are content and style, respectively.
   The fourth   column shows that
   the knowledge bases of  Users $1$ and $2$ are a large size pink letter $U$ in regular font, and a large size red letter $T$ in regular font, respectively.
The fifth column   shows  that  the reconstructed image     of User 1 is a large size pink letter $F$ in regular font, and the reconstructed data  of User 2 is a  small size red letter $T$ in italics. Thus, only the intended semantic features are transmitted to users, and the unintended semantic features are generated  based on
the knowledge base.
Moreover, the input data  of  the  last two rows is  a red left pose car. Similarly, column 3 to column 5  also   verify the effectiveness of the feature selection and feature completion  of the proposed semantic BC network.

\begin{table*}[ht]
	\caption{ Transmission comparison  over    the  semantic BC network prototype}
	\centering
	\begin{tabular}{ | c | c | c | c | c|}
		\hline
		{ } & \tabincell{c}{Transmission\\  time (ms) }  & \tabincell{c}{Compression\\ ratio }  &PSNR& \tabincell{c}{Reconstructed\\ data }  \\ \hline
		\tabincell{c}{Original\\ image}& 466.44 & 1 &100
		&\begin{minipage}[b]{0.1\columnwidth}
			\centering
			\raisebox{-.5\height}{\includegraphics[width=\linewidth]{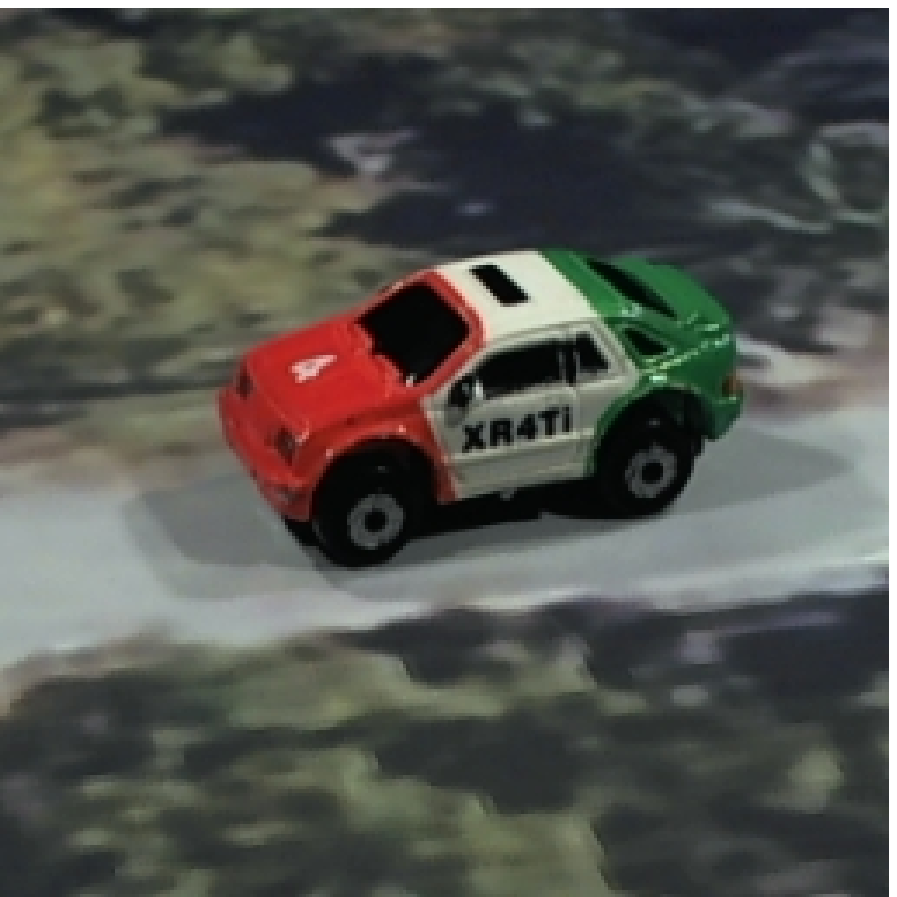}}
		\end{minipage}
		\\ \hline
		
		JPEG& 168.69& 21.22&23.72&\begin{minipage}[b]{0.1\columnwidth}
			\centering
			\raisebox{-.5\height}{\includegraphics[width=\linewidth]{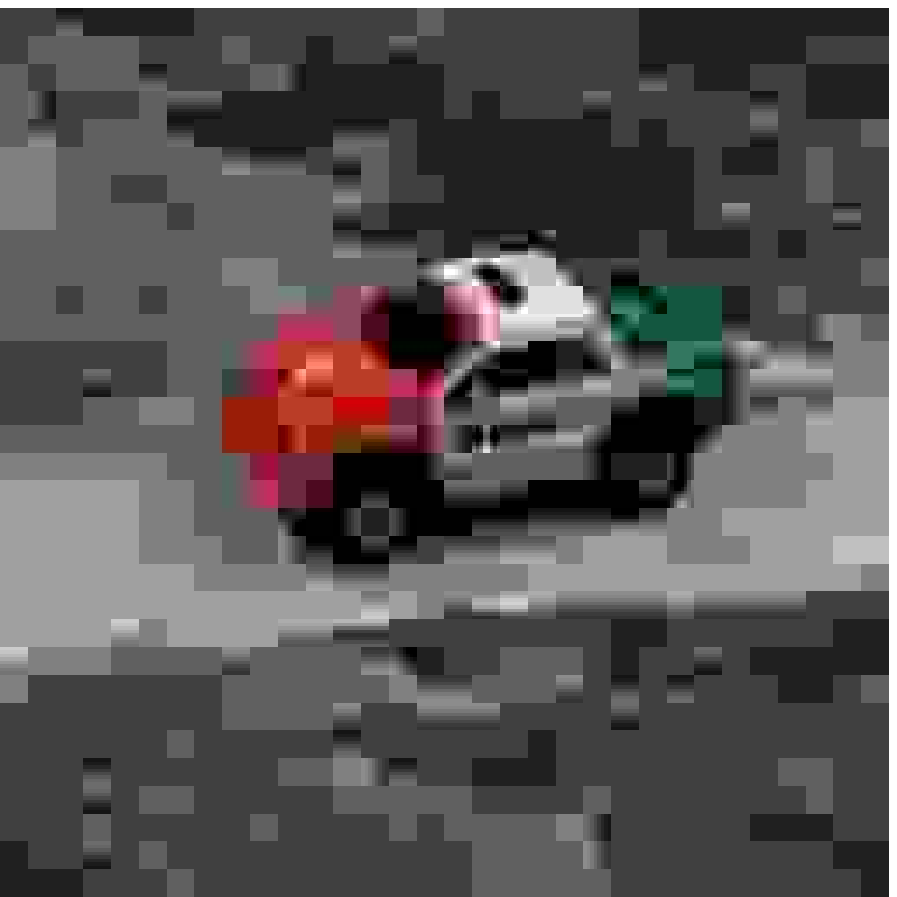}}
		\end{minipage}
		\\ \hline
		\tabincell{c}{Non-robust \\Semantic BC }  & 5 & 1966.09&27.59&\begin{minipage}[b]{0.1\columnwidth}
			\centering
			\raisebox{-.5\height}{\includegraphics[width=\linewidth]{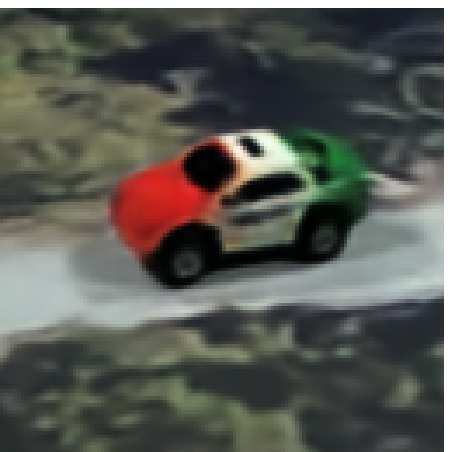}}
		\end{minipage}
		\\ \hline
		\tabincell{c}{Robust \\Semantic BC \\ ${\rm{SNR}}_{{\rm{test}}} = 8{\rm{dB}}$} & 5&1966.09&27.32&\begin{minipage}[b]{0.1\columnwidth}
			\centering
			\raisebox{-.5\height}{\includegraphics[width=\linewidth]{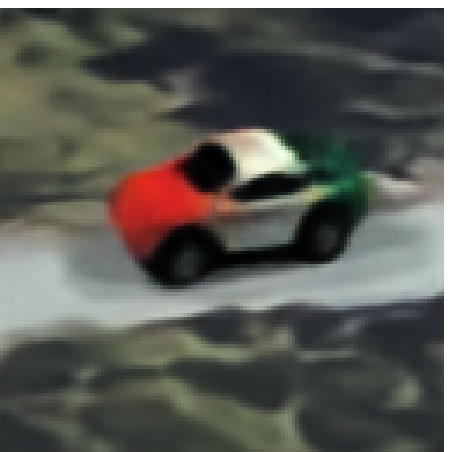}}
		\end{minipage}
		\\ \hline
		
	\end{tabular}\label{Transmission_comparison}
\end{table*}

 Finally, in Table \ref{Transmission_comparison}, we present the   transmission time (ms),    compression ratio, PSNR and reconstructed data of the original image transmission scheme,  JPEG compression scheme, non-robust semantic BC scheme, and robust semantic BC scheme with ${\rm{SNR}}_{{\rm{train}}}= 8{\rm{dB}}$.
  Table \ref{Transmission_comparison} shows that  the transmission time of    the non-robust semantic BC scheme, and robust semantic BC scheme are $5$ ms significantly lower than those of the JPEG compression scheme (168.69 ms) and the original image transmission scheme (466.44 ms).
  Moreover, the     compression ratio of the non-robust semantic BC scheme, and the robust semantic BC scheme are $1966.09$ significantly higher than those of  the JPEG compression scheme (21.22) and the original image transmission scheme.
   Therefore, the  proposed semantic BC network can significantly reduce the transmission load and transmission time.
 Moreover, the PSNRs of the robust semantic BC scheme is higher than that of the JPEG compression scheme, and is close to  that of  the non-robust BC scheme.

	%

%
%
 \section{Conclusions}

In this paper, we proposed  a practical robust features-disentangled semantic
BC framework, which can take
advantage of  the existing well-designed   standards and hardware
of bit-level communication networks.
 In our proposed framework, the semantic information was extracted and    decoupled into independent semantic features. Then, by applying features selection,
 only semantic features of interest to users are selected for transmission, and the remaining semantic features need not be sent, which   not only reduces the BC network load, but also enhances the robustness of the semantic features to channel noise.
 Moreover, we presented the optimal distortions allocation scheme for multi-source
data compression, and derived both inner and outer
bounds for the achievable rates region semantic broadcast channels.
 Furthermore, we designed a   lightweight
robust semantic BC network based on the supervised AE, and developed the corresponding hardware
proof-of-concept prototype, which is the first prototype of the semantic BC network.
   Finally, both GPU simulation and  prototype experiments demonstrated that  our proposed  semantic   BC network   is    robust to both  channel noise and channel fading, and can significantly
  improve transmission efficiency.   This paper demonstrates the viability of  a semantic BC network  design and its practical implementation.

%
%
%

%
\bibliographystyle{IEEE-unsorted}
\bibliographystyle{IEEEtran}

\bibliography{refs0611}                        

\begin{thebibliography}{10}

\bibitem{Letaief_ICM_2019}
Y.~S. J.~Z. K.~B.~Letaief~W.~Chen. and Y.~J.~A. Zhang,
\newblock ``The roadmap to 6{G}: {AI} empowered wireless networks,''
\newblock {\em IEEE Commun. Mag.}, vol.~57, no.~8, pp.~84--90, Aug. 2019.

\bibitem{Zhang_Engineering_2022}
P.~Zhang, W.~Xu, H.~Gao, K.~Niu, X.~Xu, X.~Qin, C.~Yuan, Z.~Qin, H.~Zhao,
  J.~Wei, et~al.,
\newblock ``Toward wisdom-evolutionary and primitive-concise {6G}: A new
  paradigm of semantic communication networks,''
\newblock {\em Engineering}, 2022.

\bibitem{Kountouris_CM_2021}
M.~Kountouris and N.~Pappas,
\newblock ``Semantics-empowered communication for networked intelligent
  systems,''
\newblock {\em IEEE Commun. Mag.}, vol.~59, no.~6, pp.~96--102, Jan. 2021.

\bibitem{Furuya_2009}
W.~B. M.~Furuya~R.~Sterling. and Y.~Inoue,
\newblock ``D-ila full resolution 8k projector,''
\newblock in {\em SMPTE Annual Tech Conference Expo}, 2009.

\bibitem{Cisco_2021}
Cisco,
\newblock ``Annual internet report (2018-2023) white paper,''
\newblock 2020.

\bibitem{Union_2015}
I.~Union,
\newblock ``{IMT} traffic estimates for the years 2020 to 2030,''
\newblock in {\em Report ITU}, 2015.

\bibitem{Bao_ICST_2022}
B.~Mao, F.~Tang, Y.~Kawamoto, and N.~Kato,
\newblock ``{AI} models for green communications towards {6G},''
\newblock {\em IEEE Commun. Surveys Tuts.}, vol.~24, no.~1, pp.~210--247, Nov.
  2022.

\bibitem{Niu_arXiv_2022}
K.~Niu, J.~Dai, S.~Yao, S.~Wang, Z.~Si, X.~Qin, and P.~Zhang,
\newblock ``Towards semantic communications: A paradigm shift,''
\newblock {\em arXiv preprint arXiv:2203.06692}, 2022.

\bibitem{Sana_CCNC_2022}
M.~Sana and E.~Calvanese~Strinati,
\newblock ``Learning semantics: An opportunity for effective {6G}
  communications,''
\newblock {\em arXiv preprint arXiv:2202.11958}, 2021.

\bibitem{Shi_CM_2021}
Y.~L. G.~Shi~Y.~Xiao. and X.~Xie,
\newblock ``From semantic communication to semantic-aware networking: Model,
  architecture, and open problems,''
\newblock {\em IEEE Commun. Mag.}, vol.~59, no.~8, pp.~44--50, Aug. 2021.

\bibitem{Luo_WC_2022}
X.~Luo, H.-H. Chen, and Q.~Guo,
\newblock ``Semantic communications: Overview, open issues, and future research
  directions,''
\newblock {\em IEEE Wirel. Commun.}, pp. 1--10, Jan. 2022.

\bibitem{Bao_INSW_2011}
J.~Bao, P.~Basu, M.~Dean, C.~Partridge, A.~Swami, W.~Leland, and J.~A. Hendler,
\newblock ``Towards a theory of semantic communication,''
\newblock in {\em Proc. IEEE Netw. Sci. Workshop}, pp. 110--117, Jun. 2011.

\bibitem{Yener_TCCN_2018}
A.~Y. B.~G{\"u}ler and A.~Swami,
\newblock ``The semantic communication game,''
\newblock {\em IEEE Trans. Cogn. Commun. Netw.}, vol.~4, no.~4, pp.~787--802,
  Dec. 2018.

\bibitem{Farsad_ICASSP_2018}
N.~Farsad, M.~Rao, and A.~Goldsmith,
\newblock ``Deep learning for joint source-channel coding of text,''
\newblock in {\em Proc.(ICASSP)}, pp. 2326--2330, Apr. 2018.

\bibitem{Xie_TSP_2021}
H.~Xie, Z.~Qin, L.~Geoffrey~Ye., and B.-H. Juang,
\newblock ``Deep learning enabled semantic communication systems,''
\newblock {\em IEEE Trans. Signal Process.}, vol.~69, pp.~2663--2675, Apr.
  2021.

\bibitem{Jiang_arXiv_2021}
P.~Jiang, C.-K. Wen, S.~Jin, and G.~Y. Li,
\newblock ``Deep source-channel coding for sentence semantic transmission with
  {HARQ},''
\newblock {\em arXiv preprint arXiv:2106.03009}, 2021.

\bibitem{Lu_arxiv_2021}
K.~Lu, R.~Li, X.~Chen, Z.~Zhao, and H.~Zhang,
\newblock ``Reinforcement learning-powered semantic communication via semantic
  similarity,''
\newblock {\em arXiv preprint arXiv:2108.12121}, 2021.

\bibitem{Weng_JSAC_2021}
Z.~Weng and Z.~Qin,
\newblock ``Semantic communication systems for speech transmission,''
\newblock {\em IEEE J. Sel. Areas Commun.}, vol.~39, no.~8, pp.~2434--2444,
  Aug. 2021.

\bibitem{Tong_GLOBECOM_2021}
H.~Tong, Z.~Yang, S.~Wang, Y.~Hu, W.~Saad, and C.~Yin,
\newblock ``Federated learning based audio semantic communication over wireless
  networks,''
\newblock in {\em Proc. IEEE Global Commun. Conf. (GLOBECOM)}, pp. 1--6, Feb.
  2021.

\bibitem{Shi_speech_2021}
G.~Shi, D.~Gao, X.~Song, J.~Chai, M.~Yang, X.~Xie, L.~Li, and X.~Li,
\newblock ``A new communication paradigm: from bit accuracy to semantic
  fidelity,''
\newblock {\em arXiv preprint arXiv:2101.12649}, Jan. 2021.

\bibitem{Jiang_arxiv_2022}
P.~Jiang, C.-K. Wen, S.~Jin, and G.~Y. Li,
\newblock ``Wireless semantic communications for video conferencing,''
\newblock {\em arXiv preprint arXiv:2204.07790}, 2022.

\bibitem{Tung_arxiv_2022}
T.-Y. Tung and D.~G{\"u}nd{\"u}z,
\newblock ``Deepwive: Deep-learning-aided wireless video transmission,''
\newblock {\em arXiv preprint arXiv:2111.13034}, Nov. 2021.

\bibitem{Kurka_TWC_2021}
D.~B. Kurka and D.~G{\"u}nd{\"u}z,
\newblock ``Bandwidth-agile image transmission with deep joint source-channel
  coding,''
\newblock {\em IEEE Trans. Wireless Commun.}, vol.~20, no.~12, pp.~8081--8095,
  Jun. 2021.

\bibitem{Jankowski_ISAC_2021}
M.~Jankowski, D.~G{\"u}nd{\"u}z, and K.~Mikolajczyk,
\newblock ``Wireless image retrieval at the edge,''
\newblock {\em IEEE J. Sel. Areas Commun.}, vol.~39, no.~1, pp.~89--100, Jan.
  2021.

\bibitem{Shao_JSAC_2022}
Y.~M. J.~Shao and J.~Zhang,
\newblock ``Learning task-oriented communication for edge inference: An
  information bottleneck approach,''
\newblock {\em IEEE J. Sel. Areas Commun.}, vol.~40, no.~1, pp.~197--211, Jan.
  2022.

\bibitem{Kang_arXiv_2021}
X.~Kang, B.~Song, J.~Guo, Z.~Qin, and F.~R. Yu,
\newblock ``Task-oriented image transmission for scene classification in
  unmanned aerial systems,''
\newblock {\em arXiv preprint arXiv:2112.10948}, Dec. 2021.

\bibitem{Hu_arXiv_2022}
Q.~Hu, G.~Zhang, Z.~Qin, Y.~Cai, and G.~Yu,
\newblock ``Robust semantic communications against semantic noise,''
\newblock {\em arXiv preprint arXiv:2202.03338}, Feb. 2022.

\bibitem{Huang_GLOBECOM_2021}
D.~Huang, X.~Tao, F.~Gao, and J.~Lu,
\newblock ``Deep learning-based image semantic coding for semantic
  communications,''
\newblock in {\em 2021 IEEE Global Communications Conference (GLOBECOM)}, pp.
  1--6, Dec. 2021.

\bibitem{Bourtsoulatze_TCCN_2019}
E.~Bourtsoulatze, D.~B. Kurka, and D.~G{\"u}nd{\"u}z,
\newblock ``Deep joint source-channel coding for wireless image transmission,''
\newblock {\em IEEE Trans. Cognit.Commun. Netw.}, vol.~5, no.~3, pp.~567--579,
  May. 2019.

\bibitem{Xu_TCSVT_2021}
J.~Xu, B.~Ai, W.~Chen, A.~Yang, P.~Sun, and M.~Rodrigues,
\newblock ``Wireless image transmission using deep source channel coding with
  attention modules,''
\newblock {\em IEEE Trans. Circuits Syst. Video Technol.}, May. 2021.

\bibitem{Kurika_JSAC_2020}
D.~B. Kurka and D.~G{\"u}nd{\"u}z,
\newblock ``Deepjscc-f: Deep joint source-channel coding of images with
  feedback,''
\newblock {\em IEEE J. Sel. Areas Inf. Theory}, vol.~1, no.~1, pp.~178--193,
  Apr. 2020.

\bibitem{Yang_TCCN_2022}
M.~Yang, C.~Bian, and H.-S. Kim,
\newblock ``{OFDM}-guided deep joint source channel coding for wireless
  multipath fading channels,''
\newblock {\em IEEE Trans. Cognit. Commun. Netw.}, Feb. 2022.

\bibitem{Yang_arXiv_2021}
M.~Yang and H.-S. Kim,
\newblock ``Deep joint source-channel coding for wireless image transmission
  with adaptive rate control,''
\newblock {\em arXiv preprint arXiv:2110.04456}, 2021.

\bibitem{Ding_ICASSP_2021}
M.~Ding, J.~Li, M.~Ma, and X.~Fan,
\newblock ``{SNR}-adaptive deep joint source-channel coding for wireless image
  transmission,''
\newblock in {\em Proc. IEEE Int. Conf. Acoust., Speech, Signal
  Process.(ICASSP)}, pp. 1555--1559, May. 2021.

\bibitem{Blau_ICML_2019}
Y.~Blau and T.~Michaeli,
\newblock ``Rethinking lossy compression: The rate-distortion-perception
  tradeoff,''
\newblock in {\em International Conference on Machine Learning}. PMLR, pp.
  675--685, 2019.

\bibitem{Ge_ICLR_2021}
G.~X. L.~I. Yunhao~Ge~Sami Abu-El-Haija.,
\newblock ``Zero-shot synthesis with group-supervised learning,''
\newblock {\em Proc. ICLR}, pp. 1--16, 2021.

\bibitem{Cover_Book}
T.~M. Cover and J.~A. Thomas,
\newblock {\em Elements of information theory, 2nd ed.},
\newblock New York, NY, USA: Wiley, 2006.

\bibitem{Ihara}
S.~Ihara,
\newblock ``On the capacity of channels with additive {non-Gaussian} noise,''
\newblock {\em Inform. Contr.}, vol.~37, no.~1, pp.~34--39, Sep. 1978.

\bibitem{Diederik_2014}
J.~B. Diederik P.~Kingma,
\newblock ``Adam: a method for stochastic optimization,''
\newblock {\em arXiv preprint arXiv:1412.6980}, 2014.

\bibitem{Borji_CVPR}
A.~Borji, S.~Izadi, and L.~Itti,
\newblock ``ilab-20m: A large-scale controlled object dataset to investigate
  deep learning,''
\newblock in {\em 2016 IEEE Conference on Computer Vision and Pattern
  Recognition (CVPR)}, pp. 2221--2230, 2016.

\end{thebibliography}

\end{document}